\documentclass[preprint,12pt]{emulateapj}
\usepackage{color}
\usepackage{natbib}
\usepackage{amssymb,amsmath}
\usepackage{graphicx}
\usepackage{multirow}
\usepackage[pdftex,                
bookmarks=true,                    
bookmarksnumbered=true,            
colorlinks=true,            
citecolor=blue,         
linkcolor=blue,     
menucolor=blue,         
urlcolor=cyan,       
linkbordercolor={0 0 1},           
pdfborder= {0 0 1},
frenchlinks=True]{hyperref}

\providecommand{\eprint}[1]{\href{http://arxiv.org/abs/#1}{#1}}
\providecommand{\adsurl}[1]{\href{#1}{ADS}}

\newcommand{\wtb}{WASP-12b}
\newcommand{\wt}{WASP-12}

\newcommand{\rsun}{\ensuremath{R_\odot}}

\newcommand{\hdoneb}{HD~189733b}

\newcommand{\fig}[5]{
        \begin{figure}[!bt]
        \begin{center}
        \includegraphics[#3]{#2}
      \end{center}
      
        \renewcommand{\baselinestretch}{1}
        \vspace*{-.3in}
        \caption[#4]{#5}
        \label{fig:#1}
        \end{figure}}

\def\aap{{A\&A}}		
\def\apj{{ApJ}}			
\def\apjl{{ApJ}}		
\def\apjs{{ApJS}}		
\def\pasp{{PASP}}		

\def\mnras{{MNRAS}}
\def\nat{{Nature}}

\def\methane{\ensuremath{\textrm{CH}_4}}

\def\deg{\ensuremath{^{\circ}}}

\newcommand{\name}{\mbox{Bergfors-6}}

\long\def\symbolfootnote[#1]#2{\begingroup%
  \def\thefootnote{\fnsymbol{footnote}}\footnote[#1]{#2}\endgroup} 

\shortauthors{Crossfield et al.}
\shorttitle{Re-evaluating \wtb}

\def\uedepth{\ensuremath{0.05\,\%}}
\def\edepth{\ensuremath{0.41\,\% \pm \uedepth}}

\def\edepthcorr{\ensuremath{0.45\,\% \pm 0.06\,\%}}
\def\photrms{\ensuremath{0.232\%}}
\def\edepthb{\ensuremath{0.53\,\% \pm 0.05\,\%}}
\def\etoffset{\ensuremath{7 \pm 110 \,\textrm{s} }} 
\def\eduration{\ensuremath{179.6 \pm 4.5\,\textrm{min} }}
\def\emid{\ensuremath{2455910.9090 \pm 0.0013}} 
\def\esinomega{\ensuremath{0.011 \pm 0.013}}
\def\ecosomega{\ensuremath{+0.00006 \pm 0.00091}}

\def\curlykULnoe{\ensuremath{+0.59 \textrm{~km~s}^{-1}}} 

\def\btemp{\ensuremath{3640\textrm{~K} \pm 230\textrm{~K}}}

\def\subarusep{\ensuremath{1.055'' \pm 0.026''}}
\def\subarupa{\ensuremath{250\deg \pm 1\deg}}
\def\subarufluxratio{\ensuremath{0.108 \pm 0.007}} 

\def\irtfsep{\ensuremath{1.078'' \pm 0.033''}}
\def\irtfpa{\ensuremath{249.4\deg \pm 1.1\deg}}

\def\irtffluxratio{\ensuremath{0.1048 \pm 0.0059}}

\def\guidedogscale{\ensuremath{0.116'' \pm 0.004''}}
\def\guidedogrot{\ensuremath{-0.5\deg \pm 0.6\deg}}

\def\tempB{\ensuremath{3840 \pm 70\textrm{~K}}}
\def\georatio{\ensuremath{\frac{R_\textrm{B6}}{R_\textrm{W12}} \frac{d_\textrm{W12}}{d_\textrm{B6}}}}
\def\georatioB{\ensuremath{0.452 \pm 0.015}}

\def\nstempB{\ensuremath{3660^{+85}_{-60}\textrm{~K}}}
\def\nsgeoratioB{\ensuremath{0.520^{+0.027}_{-0.037}}}
\def\nsloggB{\ensuremath{5.13^{+0.38}_{-0.22}}}

\def\rvA{\ensuremath{16.5 \pm 2.6 \textrm{~km~s}^{-1}}}
\def\rvB{\ensuremath{19.7 \pm 1.3 \textrm{~km~s}^{-1}}}

\begin{document}

\title{Re-evaluating \wtb: Strong Emission at 2.315\,\micron, 
  Deeper Occultations, and an Isothermal Atmosphere}

\slugcomment{Accepted to ApJ: 2012 Oct 17}

\author{
Ian J. M. Crossfield\altaffilmark{1}$^,$\altaffilmark{2},
Travis Barman\altaffilmark{3},
Brad M. S. Hansen\altaffilmark{2},
Ichi Tanaka\altaffilmark{4},
Tadayuki Kodama\altaffilmark{4}
}

\altaffiltext{1}{Max-Planck Instit\"ut f\"ur Astronomie, K\"onigstuhl 17, D-69117, Heidelberg, Germany; ianc@mpia.de}
\altaffiltext{2}{Department of Physics \& Astronomy, University of California Los Angeles, Los Angeles, CA 90095, USA}
\altaffiltext{3}{Lowell Observatory, 1400 West Mars Hill Road, Flagstaff, AZ 86001, USA}
\altaffiltext{4}{Subaru Telescope, National Astronomical Observatory of Japan, 650 North A'ohoku Place, Hilo, HI 96720, USA}

\begin{abstract}
  We revisit the atmospheric properties of the extremely hot Jupiter
  \wtb\ in light of several new developments.  First, new narrowband
  (2.315\,\micron) secondary eclipse photometry that we present here,
  which exhibits a planet/star flux ratio of \edepthcorr,
  corresponding to a brightness temperature of \btemp; second, recent
  Spitzer/IRAC and Hubble/WFC3 observations; and third, a recently
  observed star only 1'' from \wt, which has diluted previous
  observations and which we further characterize here. We correct past
  \wtb\ eclipse measurements for the presence of this object, and we
  revisit the interpretation of \wtb's dilution-corrected emission
  spectrum. The resulting planetary emission spectrum is
  well-approximated by a blackbody, and consequently our primary
  conclusion is that the planet's infrared photosphere is nearly
  isothermal.  Thus secondary eclipse spectroscopy is relatively
  ill-suited to constrain \wtb's atmospheric abundances, and
  transmission spectroscopy may be necessary to achieve this goal.

\end{abstract}

\keywords{infrared: stars --- planetary systems --- stars:~individual
  (\wt, \name) --- stars:~multiple --- techniques: photometric ---
  techniques:~spectroscopic --- eclipses}

\section{Introduction}

Transiting extrasolar planets allow the exciting possibility of
studying the intrinsic physical properties of these planets. The
latest new frontier to emerge is the detailed study of molecular
chemistry in the atmospheres of these planets, many of which exist in
intensely irradiated environments.  Recent years have seen
rapid strides in this direction, with measurements of precise masses
and radii, detection of secondary eclipses and phase curves and the
start of ground-based spectroscopy
\citep{redfield:2008,swain:2010,bean:2010}.  Based on observed
day/night temperature contrasts \citep[e.g.,][]{cowan:2011},
atmospheric circulation patterns \citep{knutson:2009a}, and
atmospheric chemistry \citep{stevenson:2010,madhusudhan:2011} these
planets' atmospheres are likely to be quite different from anything
previously known.

\subsection{Introducing the \wt\ System}
\label{sec:w12intro}
A prime example is the transiting Hot Jupiter \wtb, which is one of
the largest and hottest transiting planets known
\citep{hebb:2009,chan:2011,maciejewski:2011}.  The planet is
significantly overinflated compared to standard interior models
\citep{fortney:2007}, though its radius and age can be explained by an
appropriate dynamical history involving an initially eccentric orbit
and subsequent interior dissipation of tidal torques
\citep{ibgui:2011}.  Radial velocity measurements associated with the
initial transit discovery and the first occultation observation both
suggested \wtb\ had a nonzero eccentricity
\citep{hebb:2009,lopez-morales:2010}.  However, subsequent orbital
characterization via timing of secondary eclipses
\citep{croll:2011a,campo:2011,cowan:2012} and further radial velocity
measurements \citep{husnoo:2011} set an upper limit on the
eccentricity of $\sim0.03 (1\sigma)$. The 2.315\,\micron\ narrow band
eclipse we present here is also consistent with a circular orbit.

Due to its close proximity to its host star the planet is thought to
be significantly distorted and may even be undergoing Roche lobe
overflow \citep{li:2010}.  Such overflow, if verified, would be the
first evidence of the tidal inflation instability
\citep{gu:2003}. Possible evidence for the overflow scenario has come
(1) from HST/COS UV spectra taken during transit \citep{fossati:2010},
which show tentative evidence of a deeper transit with earlier ingress
than observed in the optical \citep{hebb:2009}, (2) from a tentative
detection of an extended Ks band secondary eclipse duration
\citep{croll:2011a}, which could be interpreted as an opaque accretion
stream or disk, and (3) from Spitzer/IRAC phase curve observations of
\wtb, which detect ellipsoidal variations from the planet at
4.5\,\micron\ at a level consistent with a planet filling (or
overfilling) its Roche lobe \citep{cowan:2012}.  However: (1) there is
no evidence for an extended occultation duration in Spitzer/IRAC
observations \citep{campo:2011} or in the 2.315\,\micron\ narrowband
eclipse we present here; (2) degeneracies between ellipsoidal
variations, thermal phase variations, and instrumental systematics
prevent an unambiguous determination of \wtb's geometry from the
Spitzer observations \citep{cowan:2012}; and (3) recent HST/WFC3
secondary eclipse spectroscopy suggest that \wtb\ is not substantially
distorted \citep{swain:2012}.

\wtb\ is intensely irradiated by its host star, making the planet one
of the hottest known and giving it a favorable ($\gtrsim 10^{-3}$) NIR
planet/star flux contrast ratio; its atmosphere has quickly become one
of the best-studied outside the Solar System.  The planet's large
size, low density, and high temperature motivated a flurry of optical,
\citep{lopez-morales:2010}, NIR \citep{croll:2011a}, and mid-infrared
\citep{campo:2011} secondary eclipse photometry has been interpreted to reflect
an atmosphere with an unusual carbon to oxygen (C/O) ratio greater
than one \citep{madhusudhan:2011}. Subsequent ground-based
observations \citep{zhao:2012,crossfield:2012} and the recent WFC3
1.1--1.7\,\micron\ spectrum \citep{swain:2012} are consistent with
these earlier measurements and the C/O$>1$ model, but the
2.315\,\micron\ eclipse we present here is inconsistent (at
$>3\sigma$) with such models. In addition, under the so-called ``null
hypothesis'' (i.e., a spherical planet) of \cite{cowan:2012} the IRAC
4.5\,\micron\ secondary eclipse is significantly deeper than the previous
measurement \citep{campo:2011}, suggesting less absorption by CO and
weakening the case for a high C/O ratio.

As yet transmission spectroscopy \citep[which determines atmospheric
opacity at a planet's limb via multi-wavelength transit
measurements][]{seager:2000} has so far been limited for this system.
Optical transit measurements show some disagreement
\citep{hebb:2009,chan:2011,maciejewski:2011}, which makes
interpretation difficult.  Spitzer/IRAC transit observations suggest
that \wtb's radius may be greater at 3.6\,\micron\ than at
4.5\,\micron\ \citep{cowan:2012}, but only if the planet is much more
prolate ($R_\textrm{long}/R_p = 1.8$) than suggested by WFC3
observations \citep[$3\sigma$ upper limit of 1.7;][]{swain:2012}.
Under the null hypothesis of \cite{cowan:2012}, the transit radius is
larger at 4.5\,\micron\ (as expected from atmospheric models).  \wtb's
low density and high temperature ensure that this planet will continue
to be a target for future efforts in this direction; if (as we
suggest) the planet's atmosphere is in fact nearly isothermal at the
pressures probed in secondary eclipse, transmission spectroscopy may be the only
hope for constraining \wtb's atmospheric composition.

Thus significant uncertainties remain in the interpretation of the
current ensemble of atmospheric measurements. At the moment this
situation is typical even for the best-characterized systems
\citep{madhusudhan:2010} because (a) broadband photometry averages
over features caused by separate opacity sources and (b) atmospheric
models have many more free parameters than there are observational
constraints.  When properly calibrated, spectrally resolved
measurements can break some of these degeneracies. Such results can
test the interpretation of photometric observations at higher
resolution, and can more precisely refine estimates of atmospheric
abundances, constrain planetary temperature structures, and provide
deeper insight into high-temperature exoplanetary atmospheres.  These
goals provided the motivation for our earlier ground-based
spectroscopy of \wtb\ \citep{crossfield:2011} and serve as the impetus
for the analysis presented here.

\subsection{Paper Outline}
This paper presents new secondary eclipse observations of \wtb's
emission in a narrow band centered at 2.315\,\micron, our detection
and characterization of a cool star (which we call \name) with high
surface gravity near \wt, a correction of past eclipse 
measurements for the dilution caused by \name, and our interpretation
of \wtb's atmospheric emission.

We describe our secondary eclipse observations and initial data
reduction in Sec.~\ref{sec:nb_obs}.  As described in
Sec.~\ref{sec:nb_fitting} we fit numerous model light curves to the
data, select the statistically optimal combination of parameters to
use, and present the results of this eclipse. In
Sec.~\ref{sec:startwo} we describe our analysis of \name's properties,
and in Sec.~\ref{sec:revision} we use the results of this analysis to
correct past transits and occultations of \wtb. In
Sec.~\ref{sec:atmosphere} we discuss our analysis of \wtb's corrected
emission spectrum and provide updated constraints on the planet's
bolometric luminosity.  Finally, we conclude and suggest relevant
possibilities for followup in Sec.~\ref{sec:nb_conclusion}.

\section{Subaru/MOIRCS Narrowband Time-series Photometry}
\label{sec:nb_obs} 

\subsection{Summary of Observations}
We described recently the first tentative detection of emission from
\wtb\ via spectroscopy at the 3\,m NASA Infrared Telescope Facility
(IRTF) \citep{crossfield:2012}.  However, our precision was strongly
limited by chromatic and time-dependent slit losses resulting from the
use of a single, narrow (3'') slit.  We subsequently obtained time on
the Multi-Object InfraRed Camera and Spectrograph
\citep[MOIRCS;][]{ichikawa:2006,suzuki:2008} at Subaru Observatory to
conduct multi-object occultation spectroscopy of \wtb.  A coolant leak at
Subaru caused damage that prevented us from obtaining spectroscopy, so
we operated the instrument in imaging mode using a custom narrowband
filter. This filter  (NB2315) is centered at approximately
2.315\,\micron\ with a width at half maximum of 27~nm\footnote{A
  transmission profile of the NB2315 filter is available upon request
  from T.K.}, and so is very well suited to probe the strong
absorption feature predicted to lie at this wavelength by models used
to infer a high C/O ratio \citep[][their Figure~1]{madhusudhan:2011}.

We observed one secondary eclipse of \wtb\ on 14 Dec 2011 (UT).  The start of
observations was delayed by instrument problems, but we managed to
begin about half an hour before ingress and observed continuously
thereafter.  We observed at a position angle of 330\deg\ and read out
frames in correlated double sampling (CDS) mode with a constant
integration time of 21~s per frame, using a readout speed of 8 and two
dummy reads \citep[to suppress a known, variable-bias
effect;][]{katsuno:2003}. These readout parameters result in
substantial overhead penalties, and we averaged only one frame per
61~s over our 6.5~hr of observations (which cover an airmass range of
$1.6-1.02-1.3$). We recorded 388~frames in total.  Conditions were
nearly photometric, with stellar flux variations of 1-2\,\%\ apparent.

Following standard practices for high-precision photometry
\citep[e.g.,][]{demooij:2009,rogers:2009} we defocussed the telescope
to spread the starlight over more pixels, thereby increasing observing
efficiency and reducing the effect of residual flat fielding
errors. The instrumental seeing improved throughout the night, and to
avoid any substantially nonlinear detector response we added
additional defocus to the telescope several times. Because the Subaru
autoguider was inoperative we had to periodically apply manual offsets
to the telescope tracking.  The tracking was rather poor and despite
our corrections we observed image drifts as large as 1.2''
(10~pixels); however, subsequent software development at Subaru has
improved the tracking in the absence of the autoguider.  The
temperature of both detectors (as reported by the \texttt{CHIPBOX} FITS header
keywords) increased from 76.2~K to a constant 77.0~K over the first
$1.5-2$~hr.  All these instrumental trends are shown in
Figure~\ref{fig:nb_trends}, but we ultimately find that they do not
significantly affect our photometry.

\fig{nb_trends}{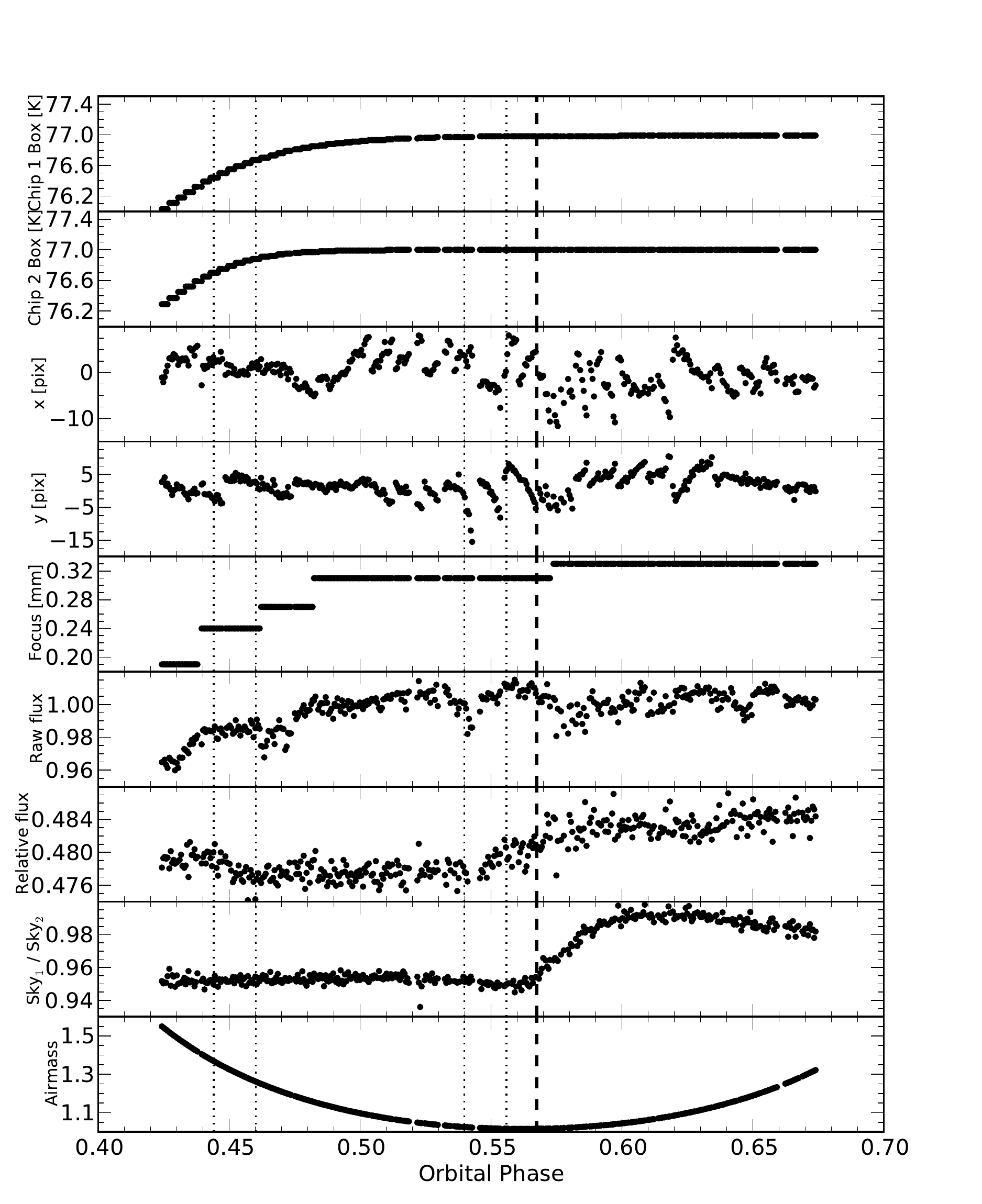}{width=3.8in}{}{Instrumental trends during
  our observations.  From top to bottom: Chip 1 and 2 electronics box
  temperatures, relative x and y motions, telescope focus encoder
  setting, \wt\ raw flux, \wt\ relative flux, ratio of median sky
  background (pre-calibration) in Chips 1 and 2, and airmass. The
  dotted lines indicate the four points of contact corresponding to  a
  circular orbit with our best-fit secondary eclipse center. The vertical dashed
  line corresponds to the onset of the anomalous trend apparent in the
  sky background and relative photometry: we exclude all data after
  this in our final analysis.}

\subsection{Initial Data Reduction}
\label{sec:nb_reduction}

MOIRCS splits its field of view across two detectors, and we reduce
the data from each detector independently. We calibrate the raw frames
following the standard MOIRCS reduction prescription, which proceeds
as follows. MOIRCS returns the UT date and time at the beginning and
end of each exposure in its FITS header.  We convert these to
BJD$_{TDB}$ for our subsequent analysis \citep{eastman:2010}.  We
dark-subtract each frame and divide the result by the stack median of
a set of dark-subtracted dome flats.  Next, we correct our data for
the intrinsic nonlinearity of infrared detector arrays
\citep{vacca:2004}\footnote{A Python implementation of our MOIRCS
  nonlinearity correction algorithm is available from the primary
  author's website.}.  Using a set of flat frames taken with typical
counts ranging from 2,000--22,000~ADU, we compute the median linearity
correction coefficients ($\mathbf{a_1}$, \ldots, $\mathbf{a_4}$) of \cite{vacca:2004}'s
Eq.~20:
\begin{equation}
\mathbf{C_{nl}} = \left( 1 + \mathbf{a_1} \mathbf{x} + 
  \mathbf{a_2} \mathbf{x}^2 + 
  \mathbf{a_3} \mathbf{x}^3 + 
  \mathbf{a_4} \mathbf{x}^4 \right)^{-1}
\end{equation} 
(where $C_{nl}$ is the ratio of an ideally linear signal to the measured signal) to be
\begin{equation*}
a_1, a_2, a_3, a_4 = 
\left( 0.00347574, -0.00436064, -0.00111471,  0.00048908 \right)
\end{equation*}
for Chip 1, and
\begin{equation*}
a_1, a_2, a_3, a_4 = 
\left( 0.0063929 , -0.0139614 ,  0.00548463, -0.00081818 \right)
\end{equation*}
for Chip 2.  We define $\mathbf{x}$ as the measured ADU counts divided
by $10^4$ to avoid very small coefficients.  We then iteratively apply
the correction algorithm in \citeauthor{vacca:2004}'s Eqs.~21-26,
while further requiring that $\mathbf{C_{nl}}$ is always $\ge
1$. Convergence typically occurs within 4-5 iterations.  The
nonlinearity correction is critical in our analysis: it changes our
relative photometry by as much as 0.5\% in some frames, and it
slightly reduces our final, residual photometric RMS from 0.234\% to
\photrms.

At this point in the analysis substantial scattered and/or background
light remains: we remove this by scaling and subtracting a
median-combined set of median-normalized, dithered, dark-subtracted
sky images.  We follow these procedures independently for data from
both channels; requiring an identical level of sky subtraction in both
channels does not significantly change our results.

We extract photometry using our own aperture photometry
package\footnote{Available from the primary author's website.}, which uses
bilinear interpolation to account for partial pixels while conserving
flux.  In each frame we extract subregions around each star, perform
1D cross-correlations to measure relative stellar motions, and
interpolate over hot pixels, stuck pixels, and any pixels more than
$6\sigma$ discrepant from their mean value.  We then recenter the
photometric apertures and perform standard aperture photometry.

In imaging mode MOIRCS offers a roughly 4'$\times$7' field of view
split equally over two $2048^2$ HAWAII-2 detectors, which allows
several comparison stars to be fit into the \wt\ field of view.  Using
stars more than about 1.8~mag fainter than \wt\ decreases our final
precision.  Our large photometric apertures also require us to avoid
choosing comparison stars with nearby companions. This leaves five
comparison stars: 2MASS stars 06302437+2937293 and 06303222+2937347
(on Chip 1) and 06302377+2939118, 06301801+2939204, and
06302280+2938338 (\wt\ is on Chip 2). Our final results are consistent
(though of lower precision) if we use fewer comparison stars or use
comparison stars falling only on a single detector. We examine the
results from photometric apertures of various sizes and ultimately use
target and inner and outer sky apertures with diameters of 39, 47, and
72~pixels (4.6'', 5.5'', and 8.4''). This choice minimizes the root
mean square (RMS) of the residuals to our model fits; the final RMS is
\photrms.

\subsection{Instrumental Systematics}
\label{sec:nbsystematics}
We plot several variable instrumental parameters, along with the 
absolute and relative photometry of \wt, in
Figure~\ref{fig:nb_trends}.  One variable dominates in terms of its
impact on our photometry: the curious trend in the ratio of the median
sky background background measured in the two detectors, which begins
an anomalous excursion as \wt\ crosses the meridian (only 10-15~min
after egress) before later stabilizing at a new level. The relative
photometry shows a qualitatively similar trend superimposed on a secondary
eclipse (visible in the raw data).

We also see this trend when dividing stellar photometry from Chip 2
(excluding \wt) by photometry from Chip 1, and we even see it (at a
lower amplitude) when comparing multiple reference stars on Chip 2
against each other; it is thus a field-dependent effect. Because the
trend begins just as \wt\ crosses the meridian, we hypothesize that
some loose component in the telescope or instrument settled in
response to the change in the direction of the gravity vector.

One possible culprit in this scenario is our narrowband filter, whose
spectral transmission profile depends on the angle of incidence of
incoming light.  To first order, increasing the angle of incidence
translates the transmission profile to shorter wavelengths.  The
filter profile intersects a particularly strong telluric absorption
(\methane) bandhead; from the vendor-supplied characterization data
for our filter we estimate that a shift in the filter's angle of
incidence of $\sim 10\deg$ could induce a photometric variation of the
magnitude observed.  However, no strong sky emission features are seen
at these wavelengths, so this scenario still has difficulty explaining
the observed variation in the sky background.  Regardless, subsequent
MOIRCS multi-object spectroscopic data do not show this anomalous
trend, a fact consistent with our hypothesis that the trend's presence
is somehow related to the NB2315 filter.

Whatever the cause of this anomalous trend, so long as we restrict our
analysis to times before orbital phase 0.5675 (the vertical dashed bar
in Figure~\ref{fig:nb_trends}) our results change by less than
$1.5\sigma$ no matter which comparison stars we choose. This choice
leaves pre- and post-eclipse baselines which are rather short. We
explored ways to use our entire data set by using the sky background
trend as a decorrelation parameter (see Section~\ref{sec:nb_fitting}
below), but such analyses resulted in larger fit residuals with
substantially higher correlations on long timescales. We therefore
proceed by excluding the later data, while acknowledging the existence
of this poorly-understood systematic effect in MOIRCS narrowband
imaging data.

\section{2.315\,\micron\ Narrowband Secondary Eclipse}
\label{sec:nb_fitting} 
We now present our analysis of the narrowband photometry discussed in
the preceding setion.  In Section~\ref{sec:fitting} we describe the
process of selecting an optimal model for our data and of fitting this
model to the data. In Section~\ref{sec:initial} we describe the
primary result of the fitting process: a 2.315\,\micron\ eclipse depth
significantly discrepant from previous predictions.  Then in
Section~\ref{sec:timing} we show that the occultation we detect has a
duration and time of center consistent with that expected for \wtb\ on
a circular orbit.

\subsection{Fitting to the Data}
\label{sec:fitting}
We fit our photometric time series with the following relation,
representing a relative secondary eclipse light curve subjected to systematic
effects:
\begin{equation}
  F_i = f_0 \left( 1 + d \ell_i \right)   \left( 1 + \sum_{j=1}^J c_{j}v_{ij} \right)  
\label{eq:fluxeqn}
\end{equation}  
The symbols are: $F_i$, the relative flux measured at timestep $i$;
$f_0$, the true relative flux; $\ell_i$, the flux in an occultation light
curve scaled to equal zero out of eclipse and $-1$ inside eclipse;
$d$, the normalized depth of secondary eclipse; $v_{ij}$, the $J$ state vectors
(i.e., image motions, sky background, airmass, orbital phase, or
low-order polynomials of these quantities) exhibiting a linearly
perturbative effect on the instrumental sensitivity; and $c_j$, the
coefficients for each state vector.  

Experience shows that the choice of instrumental model is of crucial
importance in extracting the most accurate system parameters from
transit and occultation observations \citep[e.g.,][]{campo:2011}. We
therefore explore a large region of model parameter space by fitting
our photometry using many different combinations of state vectors and
a fixed secondary eclipse time and duration. We then use the Bayesian
Information Criterion (BIC\footnote{Bayesian Information Criterion
  (BIC) = $\chi^2 + k \ln N$, where $k$ is the number of free
  parameters and $N$ the number of data points.}) to choose which of
these many models best represents our data.  To do this we first
assign uncertainties to each data point equal to the RMS of the
residuals to an eclipse fit with no additional decorrelation
variables.  We find that the BIC-minimizing model includes a transit
light curve and a linear function of time, but no additional
parameters. The model with the next-best BIC (3.2~units higher) also
include a linear function of the $x$ position of the stars on the
detector; both this model and a model including no decorrelation
parameters ($\Delta\textrm{BIC}=10.7$) return secondary eclipse depths
consistent with that of our optimal model.

We then again fit our preferred instrumental model to the data, but
now allowing the secondary eclipse center, duration, and depth to vary
while holding fixed the scaled semimajor axis ($a/R_*$) and the
orbital inclination at the values listed in Table~\ref{tab:obs}.  We
assess the uncertainties on the best-fit parameters using both the
Markov Chain Monte Carlo and prayer bead \citep[as described in ][;
see also \citeauthor{jenkins:2002} et al.~2002]{winn:2008}
approaches. The two sets of parameter distributions are quite
consistent, which suggests correlated noise does not strongly affect
our photometry.  In both cases the resulting parameter distributions
are unimodal, symmetric, approximately normal, and (excepting the
standard correlation between occultation depth and baseline flux)
uncorrelated. In the following we quote only the MCMC results, which
provide substantially denser sampling of the posterior distributions
than the prayer-bead results.

\begin{deluxetable*}{c c c c}
\tabletypesize{\scriptsize}
\tablecaption{\wtb:  2.315\,\micron\ Narrowband Secondary Eclipse Parameters \label{tab:obs}}
\tablewidth{0in}
\tablehead{
\colhead{Parameter} & \colhead{Units} & \colhead{Value} & \colhead{Reference} 
}
\startdata

$P$      & days & 1.091423  & \cite{hebb:2009}\\
$a/R_*$  & --   & 3.14   & \cite{hebb:2009} \\

$T_{c,e}$ & BJD$_{\textrm{TDB}}$ & \emid & This work  \\
$T_{\textrm{offset}}$      & s & \etoffset & This work  \\
$T_{58}$        & min &  \eduration   & This work \\
$e \cos \omega$ & --  & \ecosomega & This work\\
$e \sin \omega$ & --  & \esinomega & This work\\
$F_P/F_*$ (observed)  & -- & \edepth    & This work\\
$F_P/F_*$ (corrected) & -- &\edepthcorr & This work\\
$T_{B,2.315}$     & K  &\btemp           & This work\\

\enddata
\end{deluxetable*}

\subsection{Initial Narrowband Eclipse Depth}
\label{sec:initial}
We plot the data, our best-fit (linear baseline) model, and the
residuals in Figure~\ref{fig:bestfit}. The best-fit secondary eclipse depth is
\edepth; note that the true planet/star flux ratio is $\sim10\%$
greater than this, as we describe in Section~\ref{sec:revision} and
Table~\ref{tab:corr}.  We find shallower or deeper best-fit eclipse
depths (ranging from roughly $0.41\%$ to $0.50\%$) when using,
respectively, larger or smaller photometric apertures, which indicates
the sensitivity of this analysis to our particular choice of
parameters.  For this reason we quote the measurement uncertainty of
\uedepth\ above, which is roughly twice that predicted from our prayer-bead
analysis.

\fig{bestfit}{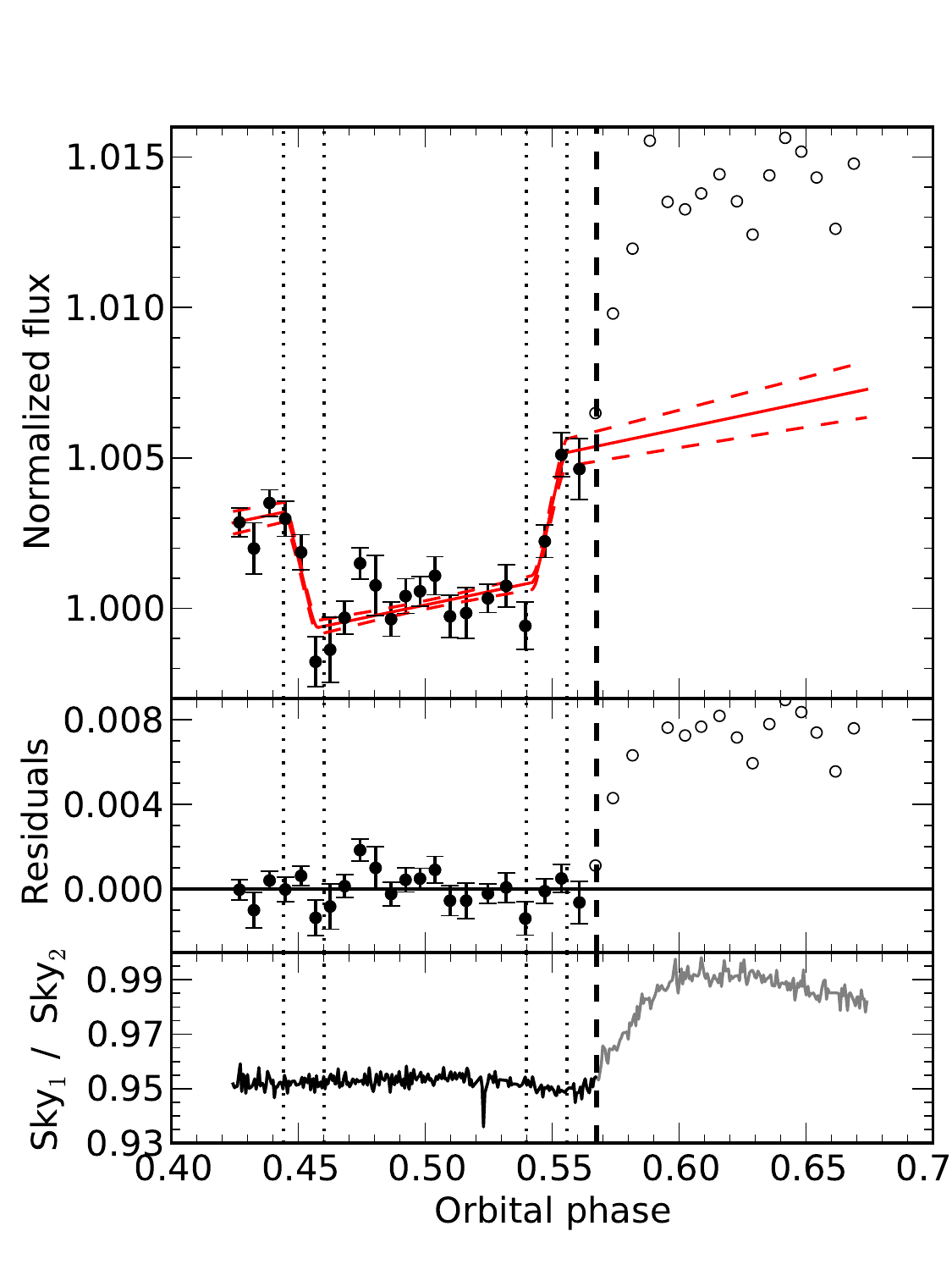}{width=3.5in}{}{Top: relative
  2.315\,\micron\ narrowband photometry of \wt\ (points, binned by a
  factor of ten for plotting purposes; errorbars are the standard
  deviation on the mean of each set of ten points) and our best-fit
  model (solid line).  The $1\sigma$ range of our model is also
  indicated by the dashed curves.  Solid points are used in our
  analysis, while open points are excluded. Middle: Residuals to the
  fit.  Bottom: differential sky background measured in the two MOIRCS
  detectors.  The vertical dashed line indicates the onset of the
  photometric ramp apparent in sky and stellar photometry; we exclude
  this data from our analysis, though the occultation depth is unchanged
  if we use all data and include the sky trend as an additional
  decorrelation parameter.  The dotted lines indicate the four points
  of contact corresponding to a circular orbit with our best-fit
  eclipse center. }

\begin{deluxetable*}{c c c c c c}
\tabletypesize{\scriptsize}
\tablecaption{Dilution Factors and Corrected \wtb\ Transit and Occultation Depths \label{tab:corr}}
\tablewidth{0pt}
\tablehead{
\colhead{Filter} & \colhead{Reported Depth} & \colhead{Aperture Fraction} & \colhead{Dilution Fraction} & \colhead{Corrected Depth} & \colhead{Reference\tablenotemark{a}} 
}
\startdata
         z & $0.00082 \pm 0.00015$ & $1.000 \pm 0.000$ & $0.0397 \pm 0.0017$ & $0.00085 \pm 0.00016$ &       LM10 \\ 
         J & $0.00131 \pm 0.00028$ & $1.000 \pm 0.000$ & $0.0606 \pm 0.0029$ & $0.00139 \pm 0.00030$ &       Cr11 \\ 
         H & $0.00176 \pm 0.00018$ & $1.000 \pm 0.000$ & $0.0830 \pm 0.0039$ & $0.00191 \pm 0.00020$ &       Cr11 \\ 
        Ks & $0.00309 \pm 0.00013$ & $1.000 \pm 0.000$ & $0.0981 \pm 0.0047$ & $0.00339 \pm 0.00014$ &       Cr11 \\ 
  Ks (MKO) & $0.00281 \pm 0.00085$ & $1.000 \pm 0.000$ & $0.0994 \pm 0.0047$ & $0.00309 \pm 0.00093$ &        Z12 \\ 
    NB2315 & $0.0041 \pm 0.0005$   & $1.000 \pm 0.000$ & $0.1002 \pm 0.0045$ & $0.0045 \pm 0.0006$ &  This work \\ 
  IRAC CH1 & $0.00379 \pm 0.00013$ & $0.902 \pm 0.018$ & $0.1168 \pm 0.0055$ & $0.00419 \pm 0.00014$ &       Ca11 \\ 
  IRAC CH2 & $0.00382 \pm 0.00019$ & $0.911 \pm 0.016$ & $0.1204 \pm 0.0060$ & $0.00424 \pm 0.00021$ &       Ca11 \\ 
  IRAC CH3 & $0.00629 \pm 0.00052$ & $0.855 \pm 0.036$ & $0.1217 \pm 0.0060$ & $0.00694 \pm 0.00057$ &       Ca11 \\ 
  IRAC CH4 & $0.00636 \pm 0.00067$ & $0.788 \pm 0.068$ & $0.1307 \pm 0.0063$ & $0.00701 \pm 0.00074$ &       Ca11 \\ 
  IRAC CH1 & $0.0033 \pm 0.0004$   & $0.850 \pm 0.038$ & $0.1168 \pm 0.0055$ & $0.00363 \pm 0.00044$ &       Co12 \\ 
  IRAC CH2 & $0.0039 \pm 0.0003$   & $0.833 \pm 0.049$ & $0.1204 \pm 0.0060$ & $0.00429 \pm 0.00033$ &       Co12 \\ 
  IRAC CH2 & $0.0050 \pm 0.0004$   & $0.833 \pm 0.049$ & $0.1204 \pm 0.0060$ & $0.00550 \pm 0.00044$ & Co12 (null)\tablenotemark{b} \\ 
\\
      V/i'\tablenotemark{c} & $0.01252 \pm 0.00045$ & $0.230 \pm 0.100$ & $0.0186 \pm 0.0023$ & $0.01257 \pm 0.00045$ &        C11\tablenotemark{c} \\ 
  IRAC CH2 & $0.0126 \pm 0.0004$   & $0.833 \pm 0.049$ & $0.1204 \pm 0.0060$ & $0.01386 \pm 0.00044$ & Co12 (null)\tablenotemark{b} \\ 
  IRAC CH1 & $0.0125 \pm 0.0003$   & $0.850 \pm 0.038$ & $0.1168 \pm 0.0055$ & $0.01374 \pm 0.00033$ &       Co12 \\ 
  IRAC CH2 & $0.0112 \pm 0.0004$   & $0.833 \pm 0.049$ & $0.1204 \pm 0.0060$ & $0.01232 \pm 0.00044$ &       Co12 \\ 
 Johnson R & $0.01380 \pm 0.00016$ & $1.000 \pm 0.000$ & $0.01571 \pm 0.00096$ & $0.01402 \pm 0.00016$ &        M11 \\ 
\\
      B    & ---                   & ---               & $0.0045 \pm 0.0015$ & --- & --- \\ 
      V    & ---                   & ---               & $0.0090 \pm 0.0030$ & --- & --- \\ 
      i'   & ---                   & ---               & $0.0281 \pm 0.0014$ & --- & --- \\ 

\enddata
\tablenotetext{a}{LM10: \cite{lopez-morales:2010}, Cr11: \cite{croll:2011a}, Z12: \cite{zhao:2012}, Ca11: \cite{campo:2011}, Co12: \cite{cowan:2012}, M11: \cite{maciejewski:2011}, C11: \cite{chan:2011}}

\tablenotetext{b}{Results from the ``null hypothesis'' of
  \cite{cowan:2012}, which assumes zero ellipsoidal variation in their
  4.5\,\micron\ observations.}

\tablenotetext{c}{These transit analyses average multiple photometric
  bands, so their correction factors may be less precise.  See Sec.~\ref{sec:revision}.  }

\end{deluxetable*}

The residuals to the best fit have a relative RMS of \photrms.  Photon
(target + sky) noise considerations predict a typical per-frame
precision of 0.06\%, so our performance is comparable to that obtained
with other NIR secondary eclipse photometry \citep[e.g.,][]{croll:2011a}.
Figure~\ref{fig:bindown} shows that our residuals bin down somewhat
more slowly than $N^{-1/2}$, indicating a moderate level of correlated
residuals.

\fig{bindown}{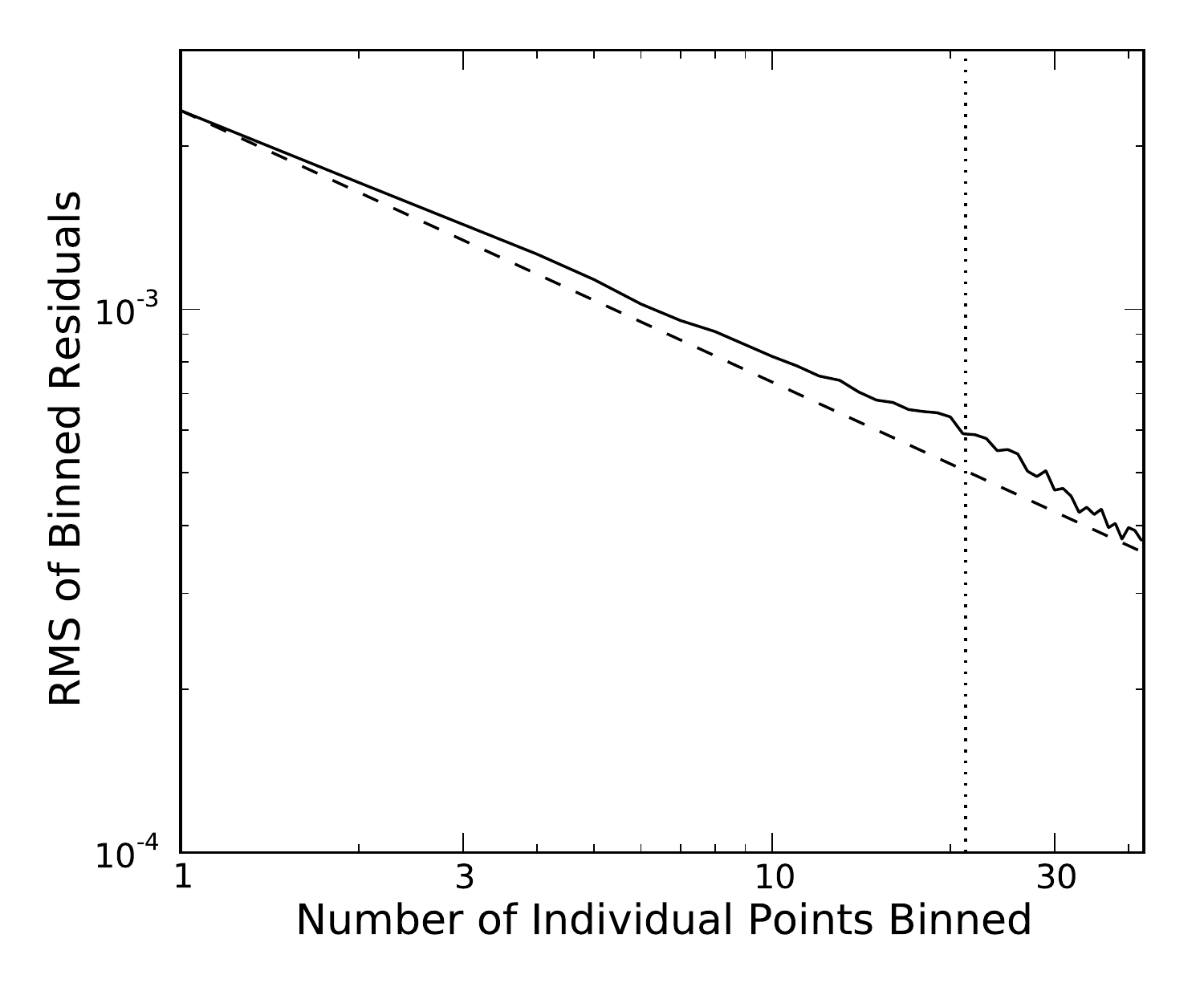}{width=3.5in}{}{RMS from
  binning the residuals shown in Figure~\ref{fig:bestfit} over
  increasing numbers of data points (solid black line). This curve is
  exceeds the $N^{-1/2}$ expectation from white noise (dashed line) by
  increasing amounts on successively longer timescales. The timescale
  of ingress or egress is indicated by the vertical dotted line. }

The largest residuals occur at a ``bump'' in the light curve at
orbital phase 0.47--0.48.  One possible explanation is that this
excursion is caused by telluric variations that are not entirely
common mode across the MOIRCS field of view. An alternate explanation
would be a short-term flare from \name.  We apply Difference Image
Analysis \citep{bramich:2008}\footnote{Our Python implementation of
  this algorithm is available at the primary author's website.} to two
images generated by co-adding individual frames during and immediately after the
bump.  The difference image shows no flux excess at the location of
\name\ during this bump, indicating that the residual feature in the
time series in not associated with \name. More exotic explanations,
such as an attribution of this bump to accretion onto \wt, should be
treated with skepticism at present.

We also attempted an alternative analysis in which we used all the
data (including that affected by the anomalous background trend) and
included the sky background trend as an additional decorrelation state
vector.  This analysis gives a marginally deeper secondary (\edepthb),
but has a higher residual RMS (0.261\%), exhibits substantially higher
levels of correlated noise when averaging on long time scales, and
shows a significantly nonzero $e \cos \omega$ \citep[inconsistent with
previous analysis; cf.][]{campo:2011,croll:2011a, cowan:2012}. These
results suggest that our simpler model, which excludes the latter part
of our data set, gives the more reliable occultation measurement.

\subsection{Occultation Duration and Timing: No Surprises}
\label{sec:timing}
We find no evidence for significant deviations in secondary eclipse
duration or in the eclipse's time of center as compared to
expectations from transit observations and a circular orbit.  We find
a best-fit eclipse duration of \eduration, and the eclipse occurs
later than predicted by \etoffset\ (after accounting for the 23~s
light travel time across the system).  Again, the parameter
distributions are unimodal and approximately normal. An analysis of
previous Ks~band observations reported a marginally longer secondary
eclipse duration \citep[$195 \pm 7$~min;][]{croll:2011a}, while a
weighted mean of Spitzer/IRAC occultations give a duration of $177.7
\pm 2.1$~min \citep[][; Cowan et al. do not report the durations of
their transits and eclipses]{campo:2011} and the z' occultation showed
a duration of 169~min \citep[with no uncertainty
reported;][]{lopez-morales:2010}.  Our narrowband measurement is
consistent with these last two values (and with the duration expected
from a circular orbit) and is within $3\sigma$ of the Ks~band
broadband results.  Our data provide no evidence for an offset or
longer-duration eclipse.

Together, the secondary eclipse timing and duration tightly constrain
the orbital eccentricity and longitude of periastron
\citep{winn:2010,seager:2011}.  We determine $e \cos \omega$ and $e
\sin \omega$ to be \ecosomega\ and \esinomega, respectively, which we
interpret as being consistent with a circular orbit and with previous
results based on eclipse and radial velocity observations
\citep{campo:2011,croll:2011a,husnoo:2011}.  The time of eclipse also
constrains the planetary velocity offset expected at transit center,
which can mimic wind-induced velocity offsets measured with
high-resolution spectroscopy
\citep{snellen:2010,fortney:2010,montalto:2011,kempton:2012}. Using
Eq.~3 of \cite{montalto:2011} we  set a $3\sigma$ upper limit
on any such orbit-induced velocity offset of \curlykULnoe.

\section{\name: An Object Very Close to \wt}
\label{sec:startwo}

\subsection{Introducing \name}
\label{sec:discx}
Before undertaking an analysis of \wtb's atmospheric properties, we
first pause to describe our improved characterization of a recently
detected point source $\sim3$~mag fainter than \wt\ and only 1'' away
\citep{bergfors:2011,bergfors:2012}.  This object requires us to revise upward past
measurements of the planet's transits and occultations.

During our Subaru observations one of us (I.T.) noticed a slight
elongation in our (defocussed) images.  This motivated us to refocus
the system at the end of the night, and we recorded the image shown in
Figure~\ref{fig:astrometry}. It clearly shows a point source roughly
1'' from \wt.  A subsequent literature search revealed that this
object was recently discovered using {\em i} and {\em z} photometry
and assigned a preliminary spectral type of K4-M1~V
\citep{bergfors:2011,bergfors:2012}. We refer to this object as \name, because it is
the sixth object  in Table~2 of \cite{bergfors:2012}.  The existence of
a bound companion at this projected separation ($\lesssim 300$~AU)
would have potentially profound implications for the dynamical history
of the system, and could provide a mechanism for Kozai-induced
eccentricity and subsequent tidal heating to inflate the planet's
radius to its present size
\citep{fabrycky:2007,nagasawa:2008,ibgui:2011}.

\fig{astrometry}{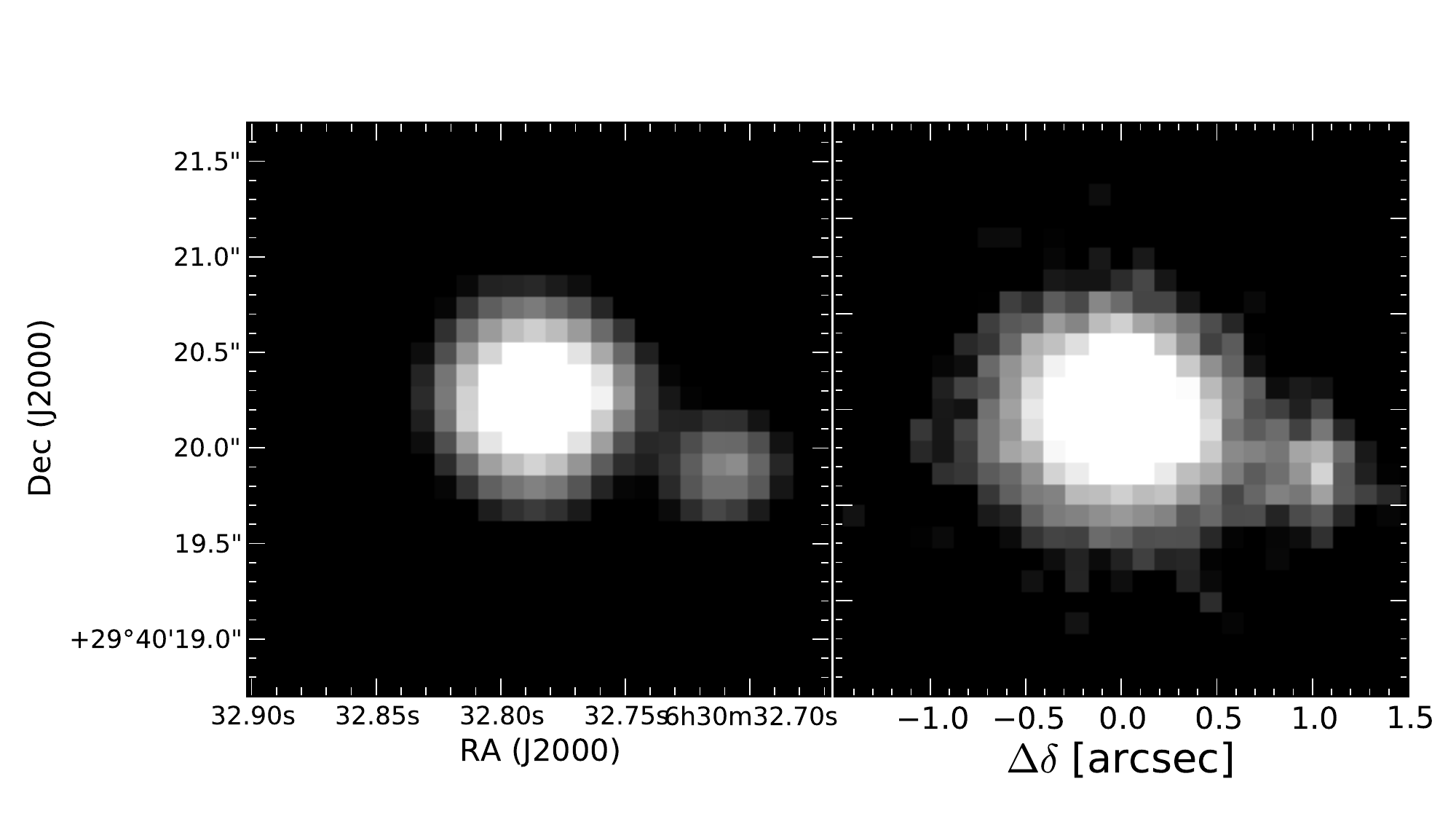}{width=3.5in}{}{Images used for
  astrometric and relative flux measurements: at left, seeing-limited
  image from Subaru/MOIRCS (2.315\,\micron\ narrowband); at right,
  speckle image from IRTF/SpeX (K$_{MKO}$).  Astrometric parameters
  derived from these images are listed in
  Table~\ref{tab:astrometry}. Both images are displayed at the same
  orientation and scale; they have different (logarithmic) color
  stretches in order to highlight the fainter companion. }


\name\ has not been remarked upon in previous optical and infrared
transit and occultation observations of the \wt\ system
\citep[][]{hebb:2009, lopez-morales:2010, chan:2011, maciejewski:2011,
  campo:2011, croll:2011a, cowan:2012,
  crossfield:2012,zhao:2012}. This is likely because with
seeing-limited or Spitzer/IRAC resolution the two objects are at best
only marginally resolved.  The case is worse for most high-precision
ground-based photometry because of the common practice of
substantially defocusing the telescope, which will clearly
preclude detection of objects such as \name.

This star falls within the photometric apertures used in most previous
analyses and dilutes the transit and secondary eclipse signals that
have been measured \citep[e.g.][]{daemgen:2009}.  In this section we
confirm the previous detection of \name\ and more tightly constrain
its spectral type.  In the following section we then correct previous
transit and occultation measurements for the photometric contamination
of \wt\ by \name.

\subsection{Observations of \name}
\subsubsection{Subaru/MOIRCS NB2315 Image}

As described above, we recorded the single well-focused MOIRCS image
shown in Figure~\ref{fig:astrometry}. We register the image's
coordinate system using the 2MASS point source catalogue
\citep{skrutskie:2006} and confirm the MOIRCS plate scale to be $0.117
\pm 0.001$''~pix$^{-1}$ (as listed in the instrument documentation).
We conservatively adopt an uncertainty of 1\deg\ in the instrumental
position angle.

\name\ sits in the wings of the \wt\ point spread function (PSF), so we
perform a simultaneous fit to the two-dimensional PSFs of both stars.
We use multiple elliptical Gaussian functions, holding the rotation
and dispersion parameters fixed in each of the model PSFs and allowing
only a single central location for each star.  Thus for $n$ Gaussian
functions we have $(5 + 5n)$ free parameters.  We set the pixel
uncertainties equal to the expectation from photon and read noise.

We find that three elliptical Gaussians minimize the BIC, so we adopt
this model and use Markov Chain Monte-Carlo (MCMC) techniques to
explore the range of valid parameter space.  We find the subsequent
parameter distributions to be unimodal and approximately Gaussian.  To
conservatively account for the uncertainties inherent in estimating
accurate photometry and astrometry from a single frame, we inflate the
parameter uncertainties estimated from our MCMC by a factor of two.
Our final determination of the flux ratio, separation, and system
position angle from the MOIRCS data are listed in
Table~\ref{tab:astrometry}. The astrometry is consistent with the
initial discovery values \citep{bergfors:2011,bergfors:2012}. We confirm this
2.315\,\micron\ flux ratio by comparing aperture photometry of \wt\
and (after subtraction of the best-fit \wt\ PSF model) of \name: this
approach gives a consistent result. The measurements presented here
are consistent with, but more precise than, estimates derived from
several of our more poorly focused MOIRCS frames.

\begin{deluxetable}{c c c c}
\tabletypesize{\scriptsize}
\tablecaption{WASP12/\name\ Astrometry \label{tab:astrometry}}
\tablewidth{0pt}
\tablehead{
\colhead{Parameter} & \colhead{Units} &  \colhead{Subaru/MOIRCS}  & \colhead{IRTF/SpeX} 
}
\startdata
Filter         & ---    &    NB2315     & K$_{MKO}$  \\
Flux ratio     & ---    &  \subarufluxratio & \irtffluxratio \\
Separation     & arcsec &  \subarusep   & \irtfsep \\
Position Angle & deg    &  \subarupa    & \irtfpa \\
Date           & UT     &  2011 Dec 14  & 2012 Feb 25 \\
\enddata
\end{deluxetable}

\subsubsection{IRTF/SpeX K Band Lucky Imaging}
\label{sec:lucky}
The well-focused Subaru image described above motivated us to acquire
additional images of \name.  On 2012 Feb 25 (UT) we imaged the \wt\
system with the IRTF/SpeX guide camera \citep{rayner:2003}, which uses
a $512^2$ Aladdin~2 Insb array with a plate scale of
0.1185''~pix$^{-1}$.  Observing through sometimes patchy clouds, we
acquired 1,200 0.4~s K$_{MKO}$ \citep{tokunaga:2002} frames (from
airmass $1.2-1.5$) and 900 0.21~s J$_{MKO}$ frames (from airmass
$1.5-2.2$).  In all observations we held the SpeX instrument rotator
at a position angle of 90\deg. The J band data were not sufficient to
reliably detect \name\ (presumably because of the increased noise
penalties resulting from the use of very short exposures and the
smaller total integration time); hereafter we discuss only the K
band data.

We calibrate the SpeX images using a median stack of internal
(thermal) flat fields and subsequently perform binlinear interpolation
over a few noticeably bad pixels. We select the top 10\% of all frames
on the basis of the peak pixel flux near the location of \wt, then use
the ``shift and add'' algorithm to align and stack these frames
\citep[our S/N is too low for more advanced
algorithms;][]{jefferies:1993, schoedel:2011}.  Changing the fraction
of frames used in our analysis from 5\% to 40\% leaves our results
unchanged within our estimated uncertainties.  (However, our most
selective analyses (which use only $1-2\%$ of the data) show a hint of
north-south elongation $\lesssim 0.3''$.  Though the resulting image
is quite noisy, we note that the initial discovery image
\citep{bergfors:2011,bergfors:2012} also shows a similar elongation. We recommend
additional high-resolution imaging of \name\ to test this elongation.)  The
final image from our standard (10\%) analysis is shown in
Figure~\ref{fig:astrometry}: in this image \wt\ exhibits an
axisymmetric PSF with a Strehl ratio of roughly 8\% and a full width
at half maximum of 0.33'' (roughly a factor of 3 better than the
seeing-limited resolution).

For astrometric reference we observed three known multiple systems
taken from Version~2012-02-12 of the Washington Visual Double Star
Catalog \citep[WDS 06295+3414, 06051+3016, and
06508+2927][]{mason:2001} moderately near \wt, with comparable
magnitudes to \wt, and with separations of 4-12''.  We took twenty
0.5-1~s frames of each system in the same region of the detector as
our \wt\ images, and in each frame we compute the centroids of both
components using standard IRAF tasks.  From these measurements and
their dispersion we derive a SpeX guider plate scale of
\guidedogscale\ and an intrinsic field rotation (i.e., true position
angle minus measured position angle) of \guidedogrot.  We adopt these
values in our subsequent analysis and list our astrometry of the WDS
stars in Table~\ref{tab:astrocal}.

\begin{deluxetable}{c c c c c c}
\tabletypesize{\scriptsize}
\tablecaption{IRTF/SpeX Astrometric Calibrators\tablenotemark{a} \label{tab:astrocal}}
\tablewidth{0pt}
\tablehead{
\colhead{WDS identifier} & \colhead{Separation} & \colhead{Position Angle} \\
\colhead{} & \colhead{[arcsec]} & \colhead{[degrees]}
}
\startdata
06295+3414 & $ 4.25 \pm 0.14$ & $ 256.8 \pm 0.6$ \\ 
06051+3016 & $11.74 \pm 0.40$ & $ 177.3 \pm 0.6$ \\ 
06508+2927 & $ 6.60 \pm 0.22$ & $ 23.7  \pm 0.6$ \\ 
\enddata
\tablenotetext{a}{All observations were made in the MKO K band on UT
  2012 Feb 25.}
\end{deluxetable}

We determine the flux ratio of the two stars using aperture
photometry.  \name\ is located in the PSF wings of \wt, so we must
account for this contamination.  Because our PSF is quite symmetric
(though distinctly non-Gaussian) we compute an average radial profile
for \wt\ (after masking out the 90\deg\ wedge of sky directed toward
\name). We reinterpolate this one-dimensional profile into a
two-dimensional model PSF. We estimate the uncertainty of the profile
by taking the standard deviation on the mean in each annular bin, and
propagate these uncertainties along with the combined photon and read
noise.  We then subtract the \wt\ model PSF from the image and compute
partial-pixel aperture photometry at the locations of \name\ and \wt.
Residuals are still apparent near the center of \wt, so we restrict
our analysis to smaller apertures: an inner aperture radius of 2.5~pix
provides the highest S/N (and least evidence for contamination) for
\name, so we use this aperture for both systems.  Our final estimate
of the $K_{MKO}$ flux ratio is listed in Table~\ref{tab:astrometry},
and it is consistent with our narrowband MOIRCS measurement.

We measure the relative astrometry of \wt\ and \name\ by computing the
centroid of \wt\ in the speckle image, and of \name\ in the
profile-subtracted image.  We estimate the uncertainties in these
measurements by bootstrap resampling \citep{press:2002}, in
which we repeat our analysis many times using synthetic data sets,
constructed by sampling (with replacement) our original set of
1,200~images.  We list the separation and position angle derived from
the IRTF speckle data in Table~\ref{tab:astrometry}.

\subsubsection{Spitzer/IRAC Imaging}
We also examined Spitzer/IRAC subarray data \citep[3.6\,\micron\ and
4.5\,\micron, from][]{cowan:2012} to search for evidence of \name. We
performed a weighted least squares fit to each median stack of
64~subarray frames (using the pixel uncertainties provided by the IRAC
calibration pipeline, Version 18.18.0) by linearly interpolating the
appropriate $5\times$ oversampled point response
functions\footnote{Available at
  \url{http://irsa.ipac.caltech.edu/data/SPITZER/docs/irac/calibrationfiles/psfprf/}}
(PRF) to account for subpixel motions.

We do see evidence for an additional point source in the IRAC data,
located approximately 1-2 pixels west-southwest of \wt. However, we
are unable to measure precise astrometry or relative photometry with
these data for several reasons.  First, the IRAC plate scale
(1.09''~pix$^{-1}$) is comparable to the \wt/\name\ separation;
second, the IRAC PSF is undersampled at these wavelengths.
Consequently, we see clear evidence for oversubtraction in the PRF
fitting at the location of \name, so we cannot reliably determine the
system flux ratio (de-weighting the pixels closest to \name, but
offset from the \wt\ core, does not change this result).  From these
measurements we estimate a flux ratio of $>7\%$ in the two IRAC
channels, consistent with our ultimate interpretation of \name\ as a
cool stellar object.

\subsubsection{Keck/NIRSPEC Spectroscopy}
We searched online data archives for additional evidence of \name, and
found a set of high-resolution K band spectra taken with Keck/NIRSPEC,
a high-resolution, cryogenic, echelle, NIR spectrograph
\citep{mclean:1998}, on UT 2010 Apr 22 (Keck Program ID C269NS,
P.I. G.~Blake).  This data set consists of 16 four-minute integrations
of \wt\ taken using the 0.432''\,$\times$\,24'' slit.
The \wt\ observations were taken at a position angle of $\sim$73\deg
(roughly aligned with \wt\ and \name), and the seeing was sufficiently
good to distinctly resolve the two components in the spectra.

We extract our spectra using our own set of Python tools to trace the
spectra in the dark-corrected and flat-fielded NIRSPEC frames. In each
echelle order of each frame, we compute a high S/N mean spectral
profile by collapsing the trace along the dispersion direction and fit
two Gaussian functions to this profile: this provides an estimate of
the projected separation of \wt\ and \name\ in each frame.  We then
fit two Gaussian functions to each resolution element while holding
constant the positions of the two sources: the amplitude of each
Gaussian represents the flux in that wavelength element.  We then
compute weighted means from the individual extracted spectra and
estimate uncertainties by measuring the variations in each pixel,
after excluding points deviating by $>3\sigma$.

Using a high-resolution simulated telluric spectrum \citep[generated
using ATRAN;][]{lord:1992} we identify known telluric lines and
compute a best-fit dispersion function in each echelle order.
Estimating a line centroid precision of 0.5~pix, we find a cubic or
quartic polynomial minimizes the BIC of these fits. The residuals to
our dispersion solutions have RMS values $\lesssim 0.1$\,\AA and
maximum excursions of $< 0.2$\,\AA.

In the raw NIRSPEC frames the spatial axis of the slit is not aligned
with the NIRSPEC detector columns, so spectra taken at the A and B nod
positions are offset from each other.  We spline-interpolate the
spectrum in each echelle order and cross-correlate it at sub-pixel
increments with a high signal to noise (S/N) template spectrum
\citep{deming:2005}.  We construct our template by taking the
temporal average, after removing outliers, of all our spectra.  A
parabolic fit to the peak of each spectrum's cross correlation
provides the optimal offset value, and we then spline-interpolate all
the spectra to a single, common reference frame. We then combine the
resulting set of aligned spectra (excluding outlying points) and
thereby provide a set of simultaneous high-resolution spectra of both
\wt\ and \name. The spectra of \wt\ and \name\ have median S/N values
of 203 and 32 per pixel, respectively.

Because the two spectra are obtained simultaneously and the objects
are separated by only 1'' we expect the telluric signature in both
spectra to be indistinguishable.  We therefore divide the spectrum of
\name\ by that of \wt\ to remove the effect of telluric absorption.
Possible misalignment of the spectrograph slit prevent these data from
usefully constraining the absolute flux ratio of these two objects,
but the data constrain tightly \name's spectral type from the relative
strengths of individual spectral features.  The NIRSPEC spectra do not
cover the wavelengths of standard gravity indicators such as Na or Ca
lines, so our subsequent analysis focuses on the two most prominent
gravity-sensitive features covered by these data: the $^{12}$CO (2,~0)
and (4,~2) bandheads located at 2.294\,\micron\ and 2.353\,\micron\
\citep{kleinmann:1986}.  As we now describe, we find that \name\ is a
hot M dwarf.

\subsection{The Spectral Type and Nature of \name}
\label{sec:nametype}

\subsubsection{Photometric Constraints}
With our four ($i, z, K_{MKO}, K_{2315}$) relative photometric
measurements and the NIRSPEC spectrum we determine the spectral type
of \name, as described below.  First, we apply the relationship
between spectral type and absolute magnitude of \cite{kraus:2007}
using our photometry.  This relationship is for main sequence stars,
and \wt\ is 25\% larger than a zero-age main sequence star of the same
mass \citep{hebb:2009,torres:2010}.  Accounting for this, and assuming
$T_{eff} = 6300 \pm 100$~K, gives a distance modulus for \wt\ of
$\mu_\textrm{W12} = 7.7 \pm 0.2$~mag \citep[significantly nearer than
the previous estimate;][]{chan:2011}.  For \name, our estimate of the
$i - K$ color \citep[after applying the color transformations of
][]{carpenter:2001} implies a main-sequence spectral type of M0-M1
($T_{eff} = 3700 \pm 100$~K).  Assuming systematic uncertainties of
0.2~mag gives $\mu_\textrm{B6} = 7.1\pm 0.2$~mag, rather closer than
\wt\ if \name\ is still on the main sequence.  That \name\ is closer
to Earth than is \wt\ is the opposite of the trend noted by
\cite{daemgen:2009}, who uniformly estimated that their faint
companions to planet host stars were more distant than the brighter
component.

Next, we fit only the relative (\name /\wt) stellar photometry using
low-resolution stellar atmosphere models \citep{castelli:2004}.  We
interpolate to the effective temperature, [Fe/H], and surface gravity
of \wt\ and hold these values fixed in the modeling.  For \name\ we
assume a metallicity equal to that of \wt\ and allow three free
parameters: surface gravity, effective temperature, and a geometric
factor ($f=$\georatio) relating the relative sizes and heliocentric
distances of the two stars.  A standard Pythonic minimizer and an
MCMC analysis using the \texttt{emcee} affine-invariant sampler
\citep{foreman-mackey:2012} provide the desired physical parameters
and their uncertainties.  The derived parameters for \name\ are
\tempB\ (which agrees well with our previous estimate of this object's
effective temperature) and $f=$\georatioB; the covariance between
these parameters is -1.73~K.

We also considered that \name\ might be an extragalactic, rather than a
stellar, contaminant \citep{luhman:2010}.  However, a
comparison of its photometric spectral energy distribution with
low-resolution galactic spectral templates \citep{assef:2010} suggests
that this explanation is unlikely.

\subsubsection{Spectroscopic Constraints}
Spectroscopy is particularly well-suited to constrain surface gravity.
We now use both the NIRSPEC spectrum and the relative photometry
described above to constrain \name's parameters, again using the
\texttt{emcee} MCMC sampler \citep{foreman-mackey:2012}. For this
analysis we use the BT-Settl library\footnote{Available online at
  \url{http://phoenix.ens-lyon.fr/}} computed using the
\texttt{PHOENIX} atmosphere code\citep{allard:2010}.  This library
provides high-resolution model spectra across a wide range of
parameter space.  We use the so-called ``hot'' models using abundances
from \cite{asplund:2009} with no alpha enhancement.  As noted
previously, we use only the two NIRSPEC echelle orders that cover CO
bandhead features -- the wavelengths from 2.273-2.308\,\micron\ and
2.344-2.380\,\micron.

For a given set of input parameters, our spectral modeling algorithm
begins by logarithmically interpolating between BT-Settl models at the
nearest values of $T_{eff}$, $\log g$, and [M/H]. The model then
applies (a) a Doppler shift, (b) a quadratic continuum normalization
(because the absolute slope and curvature of the spectrum is unknown
owing to possible slit misalignments), and (c) a convolution with a
Gaussian kernel of specified width.  Finally, we bin (not interpolate)
the model spectrum onto our NIRSPEC pixel grid and compute relative
broadband photometry as described in the preceding section -- except
that here we propagate the uncertainties in \wt's parameters into the
modeling by performing a random draw from normal distributions in
$T_{eff}$, $\log g$, and [M/H] at each step in the MCMC analysis that
follows.

In our analysis the spectral type of \name\ is constrained almost
entirely by the $\sim2000$ spectroscopic data points, while the
geometric ratio \georatio\ is constrained only by the broadband
photometry. The results of this analysis for \name\ are an effective
temperature of \nstempB\ and a $\log g$ (cgs) of \nsloggB; we show our
best-fit model spectrum in Figure~\ref{fig:starx}.  Our derived
parameters are fully consistent with an M0 dwarf on the main sequence,
which would imply a radius of $0.5-0.6\rsun$ \citep{torres:2010}.  The
geometric ratio from our analysis is \nsgeoratioB, which implies a
radius 50\% larger than expected for a main-sequence dwarf lying at
the same distance at \wt.  Our spectral analysis therefore suggests
that \name\ lies approximately 50\% closer to Earth than does \wt\ and
that it represents a chance foreground alignment.

\fig{starx}{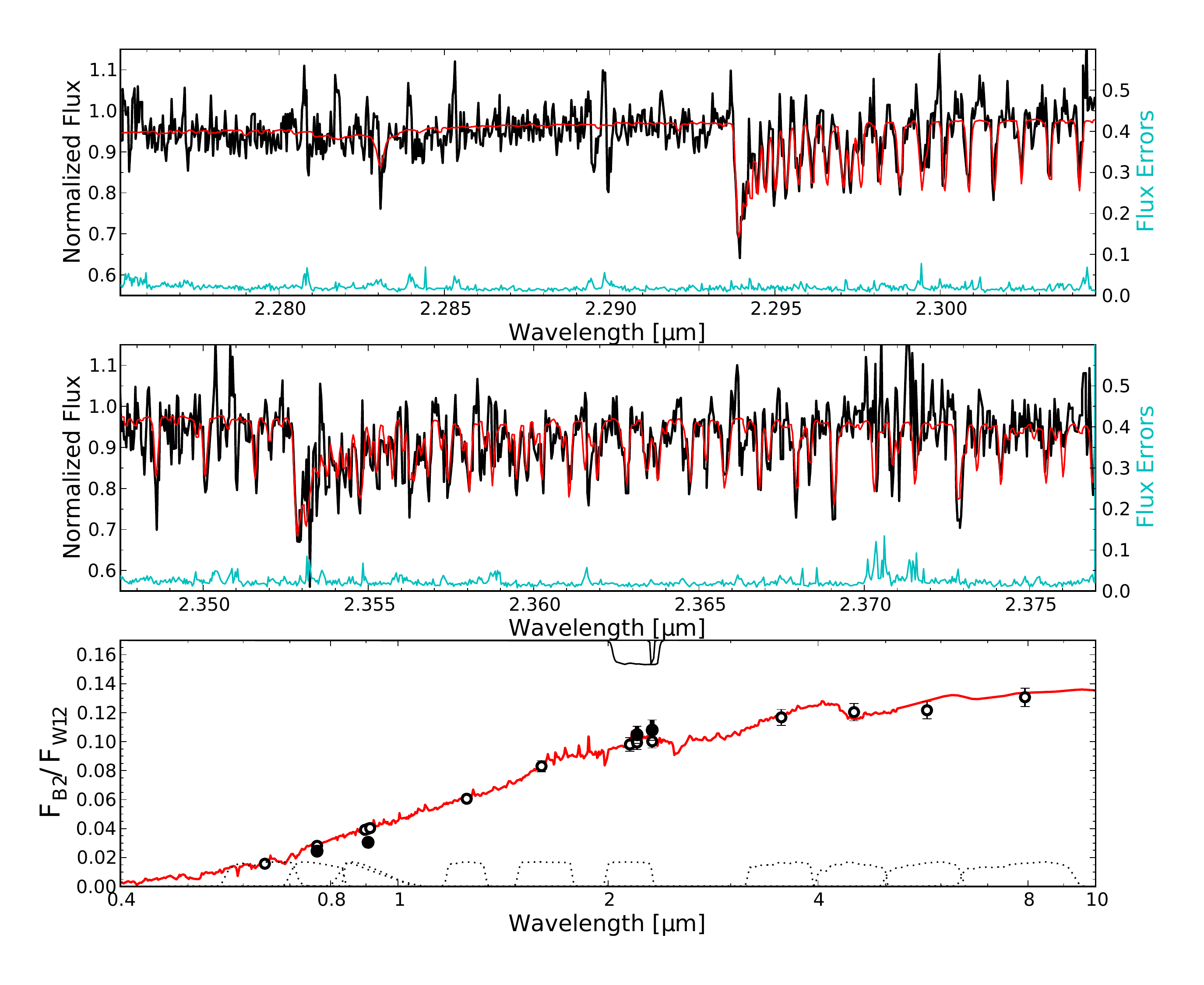}{width=3.5in}{}{
  Keck/NIRSPEC spectrum of \name\ (black curve, top and middle)
  and relative photometry of \wt\ and \name\ (filled points, bottom), 
  and our best-fit Phoenix/BT-Settl model (red curve).  As discussed in
  Section~\ref{sec:nametype}, the ensemble of measurements suggests \name\
  is a hot M dwarf located 50\% closer to Earth than \wt. 
  \textit{Top and Middle}: The blue curves show our estimated
  spectroscopic measurement uncertainties; the vertical scale for
  these curves is indicated at right.  \textit{Bottom}: The open
  points at bottom are the inferred photometric dilutions of transits
  or occultations measured in various bandpasses (tabulated in
  Table~\ref{tab:corr}); we indicate filters used in this analysis
  with solid lines, while other filters are denoted with dashed lines.
  The error bars of the open points represent the 68.3\% confidence
  intervals on these dilution estimates, taking into account the
  uncertainties in our fit.}

\subsection{Radial Velocities}
To estimate the radial velocities of the two stellar components in our
telluric-corrected \name/\wt\ ratio spectrum we cross-correlate with
BT-Settl models \citep{allard:2010}.  We cross-correlate each echelle
order of the ratio spectrum with models with effective temperatures of
6200~K and 3800~K.  We also cross-correlate the spectra (before
telluric correction) with the high-resolution atmospheric transmission
profile of \cite{hinkle:2003} to establish our observational reference
frame.

The radial velocity of \name\ is constrained mainly by the strong CO
bands lying redward of 2.29\,\micron, while \wt's is tightly
constrained only by the broad (FWHM~$=8.1$\AA) Brackett~$\gamma$ line.
After correcting for the Earth's velocity along the line of sight
(using the Python routine \texttt{astrolib.baryvel}) we estimate
radial velocities for \wt\ and \name\ of \rvA\ and \rvB, respectively.
These values are consistent with the radial velocity of
19.1~km~s$^{-1}$ derived from \wtb's initial radial velocity
measurements \citep{hebb:2009,campo:2011}. This common velocity is
consistent with a scenario in which either \wt\ and \name\ are
gravitationally bound and share a common three-dimensional space
motion, or in which the consistency of the two stars' radial
velocities is merely a coincidence.

We also find no evidence for multiple spectral line profiles, which
would indicate \name\ is an unresolved binary.  The cross-correlation
profiles of our data and spectral template have full-widths at
half-maximum of approximately 14~km~s$^{-1}$.  All three echelle
orders containing strong CO features (centered on 2.29, 2.36, and
2.44\,\micron) show unimodal cross-correlation peaks and no evidence
of the secondary peaks that would suggest an additional cool
companion.  Thus \name\ shows no spectroscopic evidence of binarity.

%
%
%
%

\subsection{Interpretation of \name}
We conclude that \name\ is a cool star showing absorption features
consistent with a high surface gravity.  Our spectroscopy implies that
\name\ is a main sequence star, and our relative photometry suggests
that \name\ lies 50\% closer to Earth than does \wt.  In this
scenario, the hint of elongation alluded to in Sec.~\ref{sec:lucky} is
spurious and the consistent radial velocities of \wt\ and \name\ is
coincidental.  However, if \name\ were a binary M dwarf system
observed near conjunction then the binary and \wt\ could lie at the
same distance from Earth; if the three components were bound this
scenario would offer a natural explanation for the consistent systemic
velocities.  Because of the intriguing possibilities inherent in a
gravitationally bound arrangement, we briefly discuss the implications
of such a scenario below.

At the distance of \wt, the projected separation of \name\ is roughly
$400 \pm 100$~AU.  Such an object would have an orbital period of
several thousand years, and as such would be marginally compatible
with the inferred upper limit of binary separation ($\sim 300$~AU)
needed to substantially influence planetary migration and dynamics
\citep{desidera:2007}.  Assuming \name\ is near apastron implies a
Kozai oscillation timescale \citep{fabrycky:2007} of (very roughly)
$20 [(1-e_\textrm{B6}) / (1 + e_\textrm{B6})]^{3/2}$~Gyr, where
$e_\textrm{B6}$ is \name's orbital eccentricity.  For substantial
Kozai interactions to have taken place during \wt's lifetime
\citep{hebb:2009,chan:2011} we thus require $e_\textrm{B6} > 0.7$.
Long-period binaries with such high eccentricities are rare, but they
do exist \citep{duquennoy:1991}.

\cite{fabrycky:2007} predict that hot Jupiter systems with an
additional, widely-separated stellar component will preferentially
exhibit misalignment between their stellar spin and planetary orbital
axes. Recent observations of \wt\ have determined that the
sky-projected angle between the spin and orbital axes is
$59^{+20}_{-15}$deg, strongly suggesting a misaligned system
\citep{albrecht:2012}.  This may be circumstantial evidence that \wt\
has migrated via Kozai interactions and that a long-period bound
companion is required \citep[e.g.,][]{ibgui:2011}.

Ultimately, high-resolution imaging can most quickly determine whether
\wt\ and \name\ truly exhibit common proper motion and are
gravitationally bound, and whether \name\ is a single or multiple
system.  Based on \wt's proper motion \citep[$\sim
8$~mas~yr$^{-1}$;][]{zacharias:2004}, speckle or seeing-limited
astrometry of the type presented here will not be sufficient for this
purpose.  However, a two-year baseline of large-aperture adaptive
optics imaging \citep{yelda:2010} could suffice to confirm or
rule out common proper motion.

\section{Revising Past Transit and Secondary Eclipses}
\label{sec:revision}
Because \name\ was not noted in previous transit and secondary eclipse
observations of \wtb, these flux diminutions were diluted by this
faint star's constant baseline flux. This effect is largest in the
infrared; the results of optical observations change by only a few
percent, less than their typical uncertainties. Although a full
re-evaluation of \wt's system parameters is beyond the scope of this
work, we correct the depth measurements for the contamination effect
and present revised transit and occultation depths below.  We
propagate the uncertainties in \name's effective temperature into our
estimates of the photometric dilution caused by \name, which we list
in Table~\ref{tab:corr}.  In some secondary eclipses, and in all
transits, the corrections we apply change the previously reported
depths by $>1\sigma$.

Essentially all the light from \name\ lies within the apertures of
ground-based observations
\citep{hebb:2009,lopez-morales:2010,croll:2011a,chan:2011,maciejewski:2011,
  zhao:2012}, but Spitzer/IRAC analyses use narrower apertures
\citep{campo:2011,cowan:2012} and so only a portion of \name's
starlight contaminates these secondary eclipse measurement.  To estimate the
IRAC contamination fraction we generate 10$\times$ super-sampled PSFs
for all four IRAC channels\footnote{Using Tiny Tim; available at
  \url{http://ssc.spitzer.caltech.edu/}}, using a 6300~K blackbody
spectrum simulated at the center of the instrument field of view.  We
then compute aperture photometry using the reported photometric
aperture diameters \citep{campo:2011,cowan:2012} at a position offset
by 1.05'' from the PSF center to estimate how much of \name's flux
fell into the \wt\ aperture in these analyses. We ignore possible
time-variable illumination caused by the intrapixel effect
\citep{charbonneau:2005}.  Note that the phase curve observations of
\cite{cowan:2012} are also diluted by \name, and must be revised
upward by the same factors as indicated in Table~\ref{tab:corr}.  This
in turn increases the ellipsoidal variation inferred from these
measurements, placing the IRAC 4.5\,\micron\ results in even stronger
conflict with those from WFC3 \citep{swain:2012}.

The correction of optical transit measurements is complicated by the
common, but deplorable, practice of reporting a single transit depth when using
observations taken in different bandpasses
\citep{hebb:2009,chan:2011}.  Atmospheric characterization via transit
observations depends on the fundamentally wavelength-dependent
planetary radius during transit, and so we recommend that future
analyses of multiple photometric data sets report the transit depths
measured in each bandpass in addition to a single, achromatic
value. In addition, such analyses are not always clear about the
relative weighting of data points from separate observations.
Nonetheless we attempt to estimate the relative weightings and derive
appropriate correction factors for prior multi-band optical transit
observations of \wt.

The analysis of \cite{maciejewski:2011} used only a single bandpass
(Johnson~R) and so their Johnson~R transit depth is the most reliable
optical transit measurement in Table~\ref{tab:corr}.  The analysis of
\cite{chan:2011} uses two transit data sets: 671 V band and 470 i'
band observations with residual RMS values of 2.0 and 1.2~mmag,
respectively.  This work uses an additional multiplicative term to
increase the per-point data uncertainties (1.48 and 1.57,
respectively).  Assuming that the statistics of the final reported
transit depth behaves similarly to a weighted mean, we estimate that
the two data sets constrain the transit parameters with roughly equal
weight.  

We use these weights to determine a weighted average of the correction
factors for each bandpass.  Such a weighted average is only a rough
approximation to the true correction factor, so the correction factors
for these analyses are somewhat less certain.  We thank the referee
for pointing out that the situation is even more muddled for the
discovery paper \citep{hebb:2009}, which uses three transit data sets:
227 B~band, 614 z'~band, and 6393 SuperWASP (roughly V band)
observations.  Because \cite{hebb:2009} do not report the residual RMS
scatter of the SuperWASP photometry, we do not attempt to determine
the relative weighting of these several transit observations and we do
not report a mean dilution-corrected transit depth for these data.

The dilution correction factors, and dilution-corrected transit and
occultation depths, are listed in Table~\ref{tab:corr}.  The optical
transit depths in particular are increased by a few percent, and the
planetary radius increases by only half this factor.  Owing to current
uncertainties in stellar properties, such an effect is smaller than
current uncertainties on WASP-12b's physical radius.

\section{Emission Spectrum and Atmospheric Properties}
\label{sec:atmosphere}
We now return out attention to the nature of \wtb's atmosphere as
constrained by its spectral energy distribution (SED).  In our narrow
bandpass we find a brightness temperature of \btemp\ by using the
model stellar spectrum described in the following section, modeling
WASP-12b's emission in our bandpass as a blackbody, and propagating
the $1\sigma$ uncertainties in our occultation measurement. This
brightness temperature is rather higher than the planet's equilibrium
temperature of $2990\pm110$~K \citep[computed using the Bond albedo
and recirculation efficiencies from ][]{cowan:2012} and is higher than
inferred in any broad photometric bandpass
\citep{lopez-morales:2010,croll:2011a,campo:2011,cowan:2012}. The
CFHT/WIRCam Ks filter cuts off just where the NB2315 filter cuts in,
so our measurement is not in conflict with this previous secondary
eclipse observation \citep{croll:2011a}.

In the following, we use the weighted averages of the two sets of IRAC
(3.6\,\micron\ and 4.5\,\micron) measurements \citep[][; see also
Table~\ref{tab:corr}]{campo:2011,cowan:2012}.  Two possible
4.5\,\micron\ secondary eclipse depths were reported by
\cite{cowan:2012}, and we use their ``null hypothesis'' consistent
with no ellipsoidal variations \citep[as implied by the constraints
placed on \wtb's shape by ][]{swain:2012}.

We first introduce the improved, model-independent constraints these
observations place on \wtb's bolometric luminosity in
Sec.~\ref{sec:bolometric}, and we then discuss our efforts to generate
a coherent model of the planet's emission spectrum in
Sec.~\ref{sec:modeling}.

\subsection{Bolometric Luminosity}
\label{sec:bolometric}
\wtb\ has one of the best-determined bolometric luminosities of any
extrasolar planet \citep{cowan:2011}. In an earlier work
\citep{crossfield:2012} we discussed the current constraints on the
bolometric luminosity of \wtb.  Here we update this analysis in light
of our dilution-corrected occultation measurements and recent Spitzer/IRAC
and Hubble/WFC3 observations \citep{cowan:2012,swain:2012}.

Current measurements now constrain the planet's dayside to have
$L_\textrm{bol} = (3.6-5.0) \times 10^{30} \textrm{erg~s}^{-1}$, where
the lower limit assumes the case of zero emission between the observed
bandpasses.  Here we have followed the same approach as our previous
calculation, but we now use the WFC3 spectrum \citep{swain:2012} from
1.1--1.485\,\micron\ \citep[replacing the J~band eclipse, but
retaining the H~band eclipse, of][]{croll:2011a}.  In this analysis our
narrowband measurement, and the controversy over the IRAC
4.5\,\micron\ measurement, affect the luminosity by only $0.1\times
10^{30} \textrm{erg~s}^{-1}$.

On the basis of thermal occultation and phase curve measurements,
\wtb's bolometric albedo has been inferred to be $A_B = 0.25\pm0.1$
\citep{cowan:2012}, which implies that the planet absorbs $(3.8 \pm
0.8) \times 10^{30} \textrm{erg~s}^{-1}$ from its host star.
Following the approach of \cite{crossfield:2012}, we find that this
value constrains the nightside luminosity to be $< 1.6 \times
10^{30} \textrm{erg~s}^{-1}$, consistent with the nightside luminosity
inferred by \cite{cowan:2012} of $0.06^{+0.12}_{-0.02} \times 10^{30}
\textrm{erg~s}^{-1}$.  This last point further assumes that the night
side (like the day side; see below) emits approximately like a
blackbody.

We are thus nearing a bolometric luminosity sufficiently
well-constrained that we can test evolutionary models of this planet.
The current uncertainty in $L_\textrm{bol}$ is dominated by
measurements at the shortest wavelengths, suggesting that occultation
observations at wavelengths $<1\,\micron$ may be the best next step
toward an even more tightly constrained bolometric luminosity.

\subsection{Atmospheric Models}
\label{sec:modeling}
The first systematic effort to retrieve \wtb's atmospheric parameters
(using infrared broadband secondary eclipse photometry) inferred a
high C/O ratio ($>1$) and ruled out any strong temperature inversion
at the pressures probed \citep[0.01-2~bar;][]{madhusudhan:2011}.  That
study published several representative models, all of which had
$\chi^2 \sim 10$ with seven measurements and $\sim10$ free parameters
(BIC$\sim 32$).  These models also predicted a strong absorption
feature (depth $\lesssim 0.2\%$) at 2.315\,\micron: our narrowband
secondary eclipse measurement rules out this absorption feature at
$>3\sigma$.  This narrowband result, our characterization of \name\
and correction for its diluting effect, and the alternative
Spitzer/IRAC 4.5\,\micron\ eclipse depth of \cite{cowan:2012} all
indicate the need for a new analysis of \wtb's atmospheric properties.

To better understand the nature of the planet's dayside emission we
constructed a variety of atmosphere models for WASP-12b, following
\cite{barman:2001,barman:2005}.  We compared model SEDs to the
corrected occultation depths from Table~\ref{tab:corr}. For this
exercise we adopt the \cite{cowan:2012} 4.5\,\micron\ value without
ellipsoidal variations (their ``null hypothesis'').  As discussed by
\citeauthor{cowan:2012}, when ellipsoidal variations are allowed, the
inferred planet elongation in this band is substantially larger than
expected.  Indeed, recent NIR HST/WFC3 observations exclude the large
ellipsoidal variations inferred from the Spitzer/IRAC analysis
\citep{swain:2012}. Furthermore, the measured primary transit radius
at 4.5\,\micron\ differs dramatically from radii at other wavelengths
as well as predicted transit spectra.  Given that two very unexpected
physical properties are required to explain the ellipsoidal
variations, the 4.5\,\micron\ secondary eclipse value without
ellipsoidal variations is likely more reliable (Cowan~2012, private
communication). Finally, our ultimate result -- that the high C/O result
is not justified by current photometric data -- does not change even
if we use the average of the \citeauthor{campo:2011} and
\citeauthor{cowan:2012} results.

By far the model that best matches the broad band occultation
photometry is a 3000~K black body (upper panel of Fig.~\ref{fig:sed},
and Fig.~\ref{fig:profiles}), with $\chi^2 = 15$ with only two free
parameters (BIC$=19$).  Taking the mean of the \citeauthor{campo:2011}
and \citeauthor{cowan:2012} results gives a black body model with
higher $\chi^2 = 25$ and BIC$=29$, but this BIC value is still smaller
than that for the high C/O models. Discussed above. This lower BIC
value indicates that current observations do not justify the use of
these more complicated models. Our best-fit black body temperature
matches the expected equilibrium temperature of the planet if
redistribution of heat to the night side is extremely inefficient.
The close match to a blackbody also indicates that the planet's
photosphere is nearly isothermal at the depths probed by these
wavelengths.

\fig{sed}{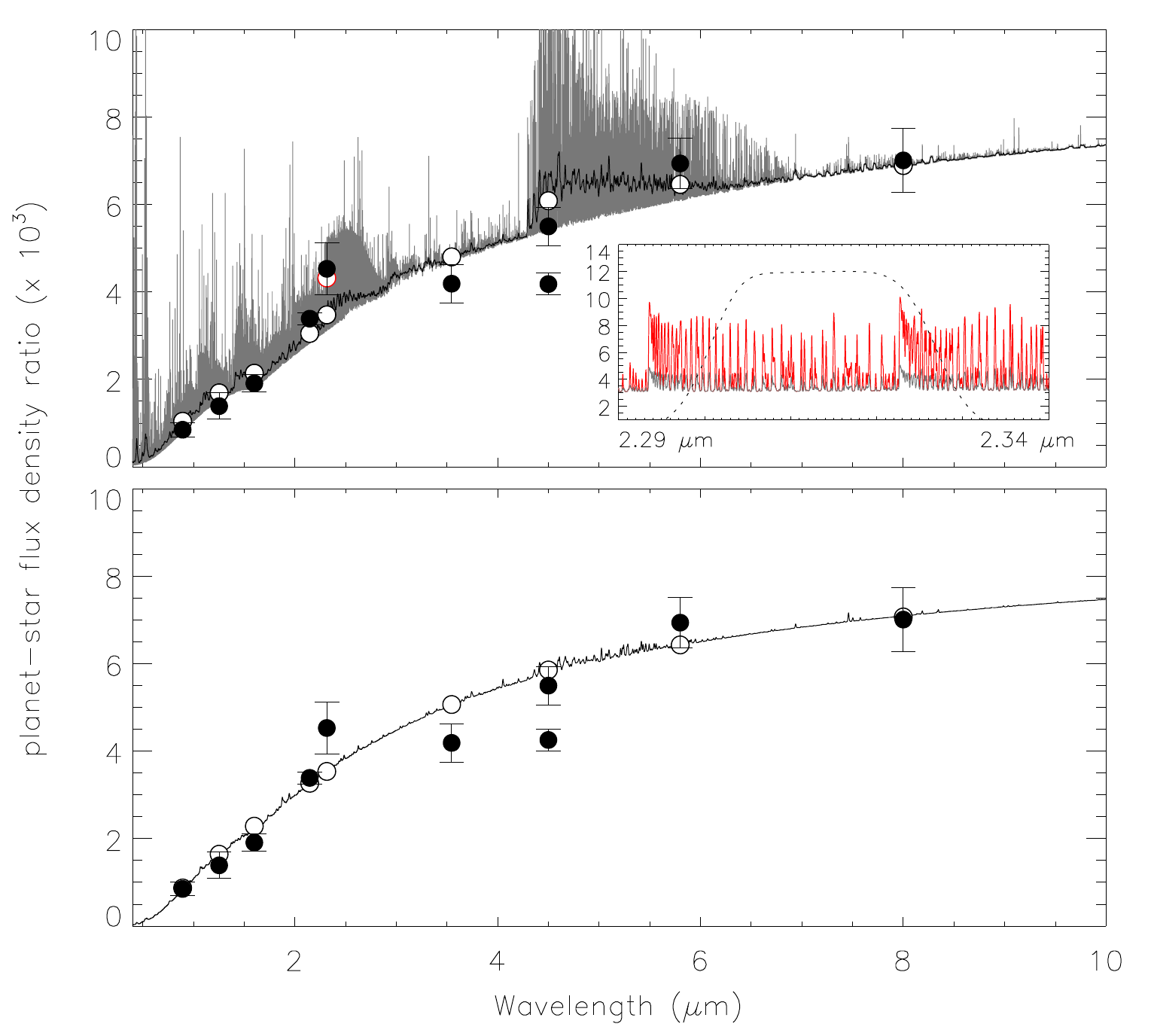}{width=3.5in}{}{WASP-12b emission
  spectrum; see Sec.~\ref{sec:atmosphere} for a full discussion. Solid
  points are the dilution-corrected photometric secondary eclipse
  depths listed in Table~\ref{tab:corr}; at 4.5\,\micron\ we plot the
  results of both \cite{campo:2011} and \cite{cowan:2012} The lower
  panel is the black body comparison (open symbols are the model,
  band-integrated, points).  The upper panel includes several spectra:
  the dark black spectrum (with open black symbols) is our solar
  abundance, $2\pi$ redistribution model spectrum (smoothed for
  plotting purposes by convolving with a Gaussian of FWHM = 100\AA).
  The grey spectrum reproduces the black model but is plotted at 100
  times higher resolution to show the many narrow emission lines
  predicted by this model.  The inset shows the narrow band region and
  a model with CO lines scaled up (red curve).  Within the inset plot,
  the ``standard'' high-res model is also shown in the same grey color
  as in the main figure.  Note the scale: these lines are narrow but
  very bright.  If such strong, narrow emission lines are present,
  they should be easily discerned by future observations. }

\fig{profiles}{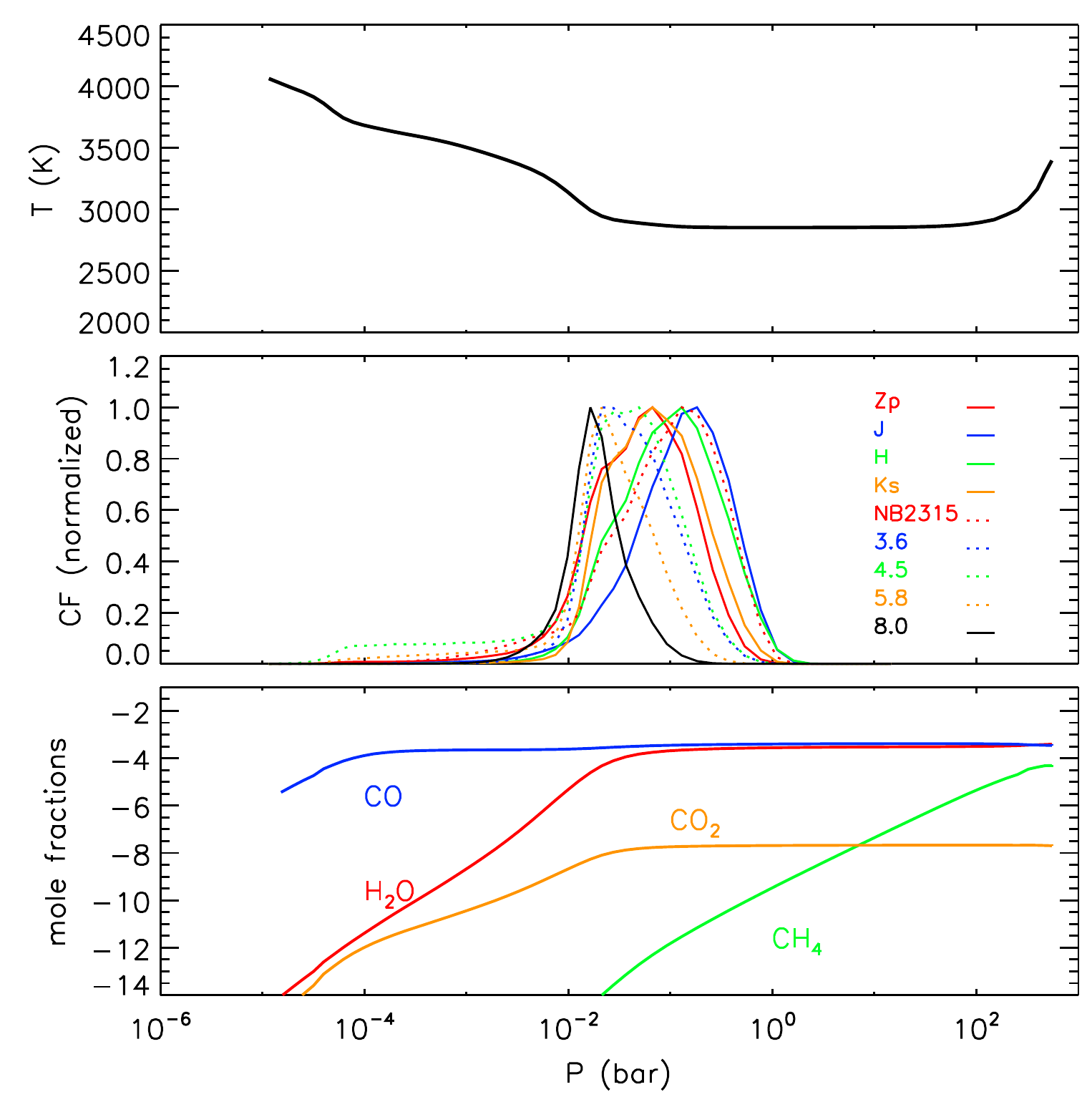}{width=3.5in}{}{Temperature-pressure
  profile (top) and molecular abundances (bottom) used to construct
  the model emission spectrum of \wtb\ shown in Fig.~\ref{fig:sed}.
  The middle panel shows the normalized contribution functions for the
  indicated filters.}

Our default irradiated atmosphere model assumes zero redistribution of
heat to the nightside (consistent with the black body analysis) and
metal abundances matching those of the host star
($\textrm{[Fe/H]}=0.3$). This model predicts a temperature inversion
above an isothermal region (Fig.~\ref{fig:profiles}); it also has
$\chi^2 = 15$; since it has substantially more free parameters than a
black body, the BIC for this model fit is substantially worse than for
the simpler model in the previous paragraph.  In such a model the
inversion extends partially across the IR photosphere resulting in
modestly inverted CO and H$_2$O bands (lower panel of
Fig.~\ref{fig:sed}). However, the model band-integrated fluxes are
similar to the black body and match the broad band observations
equally well (Fig.~\ref{fig:sed}).  Naturally, then, the blackbody
model provides a substantially lower BIC than the radiative-transfer
models \citep[both our model and the models presented by
][]{madhusudhan:2011}.

Although our narrowband measurement is fit by the simple blackbody
model (within $2\sigma$), we additionally investigated what mechanisms
might cause such a narrow band to exhibit a substantially higher
flux. In an attempt to reproduce the narrow band flux, we made a
number of ad-hoc modifications to the temperature-pressure (T-P)
profile of the default atmosphere model.  The base of the inversion
was moved to higher and lower pressures, the steepness of the
inversion was altered and several Gaussian temperature perturbations
were added.  None of these models could match the narrow-band flux
without negatively impacting the comparison at other wavelengths.  A
dozen different carbon and oxygen abundances (ranging from $7.7-9.4$
on the standard base-10 logarithmic scale, where H$_2$ abundance is
exactly 12; these abundances overlap the best-fitting regions from
Madhusudhan et al.~2011) were also explored, under the assumption of
chemical equilibrium, with similarly negative results.  Finally, we
constructed a dayside model by dividing the planet's hemisphere into
10 concentric regions centered on the substellar point.  These regions
were modeled separately, each receiving flux along the line-of-sight
to the star.  The outgoing intensities during secondary eclipse, along
the line-of-sight to the observer, were integrated to produce a
dayside spectrum following \cite{barman:2005}.  The predicted
limb-to-substellar temperature structure implies a horizontal as well
as vertical temperature inversion (i.e., along a path from the
substellar point to the terminator at a constant radius or pressure
the temperature decreases, then increases again), but the
surface-integrated fluxes from this model did not significantly differ
from the default one-dimensional model.  Our main conclusion from this
modeling exercise is that it is very unlikely that strong narrow band
emission could be reproduced by a modification to the thermal profile
alone, given the strong constraints placed on the SED by the broad
band photometry.

As discussed above, the default, 1D, model predicts a steep
temperature inversion. A direct consequence of this inversion is a
swath of narrow emission lines that form in the inverted atmospheric
layers (grey-shaded spectrum in Fig.~\ref{fig:sed}).  While the
narrow band filter does not encompass the complete CO band, it does
cover many of the predicted narrow emission lines including the strong
3-1 band.  In the default model, these lines are very bright relative
to the continuum but are too narrow to significantly impact the
photometry.  The model exercise above weakens the case for flux
increases in the pseudocontinuum, leaving the lines as a potential
explanation for the deep narrow-band secondary eclipse.  If the line fluxes are
increased by a factor of 2 to 4, the narrow band photometry can be
reproduced.  In this ad-hoc scenario, the lines would be extremely
bright compared to those of the default model (see comparison in
Fig.~\ref{fig:sed} inset).  In the default atmosphere model, CO is the
fourth most abundant molecule throughout most of the atmosphere
including much the inversion region, as shown in
Fig.~\ref{fig:profiles}. Previous studies have shown that local
thermodynamic equilibrium (LTE) is achieved for CO in the atmospheres
of isolated late-type stars \citep{ayres:1989,schweitzer:2000},
meaning that the CO lines should map the gas temperature.  However,
\cite{barman:2002} concluded that Na was well out of LTE in the upper
atmosphere of the modestly irradiated planet HD~209458b, resulting in
strong emission cores for the Na D line profiles.  Nevertheless, the
LTE assumption has not been tested for molecules in highly irradiated
atmospheres.  If such strong, narrow emission lines were present in
the atmosphere of a hot Jupiter, they should be easily discerned by
future observations.
Note that while \methane\ also has a strong band head at this
wavelength, \citep[as observed in the telluric transmission spectrum
and in spectra of L~dwarfs; cf.][]{hinkle:2003, cushing:2005}, at the
high temperature of \wtb\ atmospheric conditions would have to be much
farther from equilibrium for significant \methane\ opacity to be
apparent.  However, a careful non-LTE study is beyond the scope of
this paper and further observational confirmation of the narrow-band
emission is needed.

The claim of strong, non-LTE \methane\ emission in \hdoneb\
\citep{swain:2010}, which was based on single-slit IRTF/SpeX
spectroscopy, was disputed on the basis of contamination by telluric
effects \citep{mandell:2011}.  However, the types of spectroscopic
contamination discussed by \cite{mandell:2011} do not apply to our
narrowband relative photometry.  Changes in telluric absorption are
common mode over our narrow field of view and should be removed when
we divide the flux from \wt\ by the comparison star flux. Variable
telluric emission should be removed by the combination of global sky
frame subtraction and local background subtraction in the aperture
photometry process (Section~\ref{sec:nb_reduction}).  As
discussed in Section~\ref{sec:nbsystematics}, though the telluric
absorption feature intersecting our narrowband filter could explain
the systematic photometric ramp (seen shortly after \wt\ crosses the
meridian; Figure~\ref{fig:bestfit} and
Sec.~\ref{sec:nbsystematics}), it seems unlikely that telluric
variations could strongly corrupt the secondary eclipse depth while maintaining
such a consistent eclipse duration and time of center
(Sec.~\ref{sec:timing}).

Our analysis of the observations to date support a planet with little
to no redistribution of flux to the night side, consistent with
\cite{cowan:2012}.  We also find that photometric observations are
well-reproduced by a black body and are not yet sufficiently precise
to justify the use of more complicated models.  If WASP-12b has a
near-isothermal photosphere, then secondary eclipse data will be
poorly suited to  reveal significant compositional information.
Other highly irradiated giant planets have also been observed to host
nearly isothermal infrared photosphere, including the similarly hot
WASP-18b \citep{nymeyer:2011} and perhaps the cooler TrES-2b and
TrES-3b \citep{odonovan:2009, croll:2010,cowan:2011}.  If \wtb\ does
indeed largely radiate like a black body, previous conclusions about
composition and the the presence or absence of a temperature inversion
are significantly weakened.  If our narrow band flux measurement is
confirmed at higher precision and this flux is produced by emission
lines, there may be hope for high-resolution spectroscopic studies to
infer the planet's atmospheric composition and thermal profile
\citep[e.g.\,][]{barnes:2007a,mandell:2011,rodler:2012,brogi:2012}.

\section{Conclusions}
\label{sec:nb_conclusion} 

We have presented a deeper-than-expected secondary eclipse
(\edepthcorr) of the very hot Jupiter \wtb\ in a narrow band centered
at 2.315\,\micron.  The planet's brightness temperature at this
wavelength is \btemp, only marginally consistent with \wtb's
equilibrium temperature of $2990\pm110$~K \citep{cowan:2012}.  Our
precision is lower than expected because of an unanticipated
systematic trend affecting both sky background and stellar photometry,
but we are able to exclude data affected by this trend. The duration
and timing of the eclipse we measure from these data are consistent
with a circular orbit and with previous measurements
\citep{hebb:2009,croll:2011a,campo:2011,cowan:2012}.

Using NIR photometry and high-resolution spectroscopy, we find that
\name, a previously identified object only 1'' from \wt\
\citep{bergfors:2011,bergfors:2012}, is an M dwarf star with $T_{eff}=$\nstempB. If
this object is an unresolved binary with two components of equal mass
it could lie at the same distance from Earth as does \wt.  However,
Keck/NIRSPEC spectroscopy shows no evidence for binarity.  If single,
it likely lies closer to Earth than does \wt.  Adaptive optics imaging
on large-aperture telescopes will be necessary to conduct the proper
motion studies necessary to discriminate between these two scenarios.
If \wt\ and \name\ are gravitationally bound, further simulations
\citep[e.g., ][]{ibgui:2011} should be undertaken to determine whether
Kozai interactions with an object with \name's characteristics could
have caused \wtb's inward migration and, through tidal pumping, have
inflated the planet's radius \citep{bodenheimer:2001}.

\name\ has heretofore passed unnoticed in previous transit and
occultation analyses, which has caused the measured depths of these
past events to be underestimated. We use our constraints on \name\ to
infer and correct for the dilution of these past observations, in
several cases increasing depths by $>1\sigma$. Thus \wtb\ is rather
hotter and slightly larger (by $1-2\%$) than previously reported.
These changes emphasize the importance of high-resolution imaging
surveys in the vicinity of newly discovered transiting planets.

The ensemble of dilution-corrected secondary eclipse measurements
suggests that \wtb's atmosphere is largely isothermal across the
pressures probed by eclipse observations (Figs.~\ref{fig:sed}
and~\ref{fig:profiles}), with a photospheric temperature of roughly
3000~K.  This result implies that previous claims of a high carbon to
oxygen ratio for this planet \citep{madhusudhan:2011} are not yet
justified by the current photometric data.  Further observations of
the planet's 4.5\,\micron\ secondary eclipse depth is certainly
warranted to resolve the discrepancy between previous results at this
wavelength \citep{campo:2011,cowan:2012}, as our modeling efforts
indicate that achieving such a dramatically lower temperature in the
4.5\,\micron\ channel would require a major unexpected change in the
opacity source(s) across this bandpass. Regardless, our narrowband
measurement alone excludes the models used to infer this high C/O
ratio at $>3\sigma$. The lack of absorption at a wavelength where CO,
a dominant species in any atmospheric model, should exhibit strong
absorption is further evidence for a near-isothermal photosphere. Thus
secondary eclipse observations are ill-suited to determine \wtb's
atmospheric composition, and ultimately transmission spectroscopy may
be a more successful approach in pursuit of this goal.

\wtb\ is clearly an unusual object, and further observations are
clearly warranted.  Aside from the need for additional IRAC
4.5\,\micron\ occultation photometry as described above, any or all of
narrowband photometry (from the ground or, if available, using
HST/NICMOS), single- or multi-object spectroscopy
\citep[][]{swain:2010, bean:2010,mandell:2011, berta:2012}, or perhaps
by high-resolution phase curve spectroscopy
\citep[][]{barnes:2007a,brogi:2012,rodler:2012} could be of great
utility.  Finally, a better measurement of the planet's
three-dimensional shape is also highly desirable, especially given the
apparent disagreement between the degree of prolateness inferred by
\cite{cowan:2012} and \cite{swain:2012}.  If \wtb\ is substantially
prolate, three-dimensional models, ideally coupled with a general
circulation model of the planet's atmospheric dynamics, may also
provide important clues toward unraveling the mystery of \wtb's
atmospheric structure and composition.

\section*{Acknowledgements}
We thank N. Cowan for many fruitful discussions about the WASP-12
system, M. Swain and J. Bean for discussions about the systematic
trend apparent in our relative photometry, K.~Stevenson and
J.~Harrington for reiterating to us the importance of partial pixels
in high-precision aperture photometry, C.~Bergfors for discussions of
the object she discovered, and our anonymous referee for detailed
comments and sugestions which improved the quality of this paper.

I.C.~was supported by the UCLA Dissertation Year Fellowship and by
EACM.  B.H.~is supported by NASA through awards issued by JPL/Caltech
and the Space Telescope Science Center.  T.K.~acknowledges the
financial support by a Grant-in-Aid for the Scientific Research (No.\,
21340045) by the Japanese Ministry of Education, Culture, Sports,
Science and Technology, with which the NB2315 filter was made.  This
research has made use of the Exoplanet Orbit Database at
\url{http://www.exoplanets.org}, the Extrasolar Planet Encyclopedia
Explorer at \url{http://www.exoplanet.eu}, and free and open-source
software provided by the Python, SciPy, and Matplotlib communities.
We will gladly distribute our raw data products, or many of our
algorithms, to interested parties upon request.

Facilities used: Subaru, IRTF, Keck, Spitzer


\clearpage

\clearpage

\end{document}